\documentclass[fleqn,usenatbib,useAMS]{mnras}
\usepackage[english]{babel}

\newcommand{\kms}{\,km\,s$^{-1}$}

\usepackage{xcolor}
\usepackage{longtable}
\usepackage{graphicx}
\usepackage{amsmath}
\usepackage{amssymb}
\usepackage{multicol}
\usepackage{threeparttable}

\title[IRAS 02143] {Post-AGB candidate IRAS 02143+5852: Cepheid-like variability,
three-layer circumstellar dust envelope and spectral features}
\author[Ikonnikova  et al.]{N.P.~Ikonnikova$^{1}$\thanks{E-mail:
ikonnikova@sai.msu.ru}, M.A.~Burlak$^{1}$, A.V.~Dodin$^{1}$,
S.Yu.~Shugarov$^{1, 2}$, A.A.~Belinski$^{1}$, \newauthor
A.A.~Fedoteva$^{1}$,  A.M.~Tatarnikov$^{1, 3}$, R.J.~Rudy$^{4}$,
R.B.~Perry$^{5}$, S.G.~Zheltoukhov$^{1, 3}$, \newauthor
K.E.~Atapin$^{1}$
\\
\\
$^{1}$Sternberg Astronomical Institute, Lomonosov Moscow State University, Moscow 119234, Russia\\
$^{2}$Astronomical Institute of the Slovak Academy of Sciences, Tatranska Lomnica 05960, Slovakia\\
$^{3}$Faculty of Physics, Lomonosov Moscow State University, Moscow 119991, Russia\\
$^{4}$Kookoosint Scientific, 1530 Calle Portada, Camarillo, CA 93010, USA\\
$^{5}$Alabaster Scientific, P.O. Box 120, Irvington, VA 22480, USA}

\date{Accepted 2024. Received 2023; in original form 2023}

\pubyear{2024}

\begin{document}
\pagerange{\pageref{firstpage}--\pageref{lastpage}}

\maketitle

\label{firstpage}

\begin{abstract}

We present the results of multicolour
$UBVR_{\text{C}}I_{\text{C}}JHK$ photometry, spectroscopic
analysis and spectral energy distribution (SED) modelling for the
post-AGB candidate IRAS~02143+5852. We detected Cepheid-like light
variations with the full peak-to-peak amplitude $\Delta
V\sim0.9$\,mag and the pulsation period of about 24.9\,d. The
phased light curves appeared typical for the W~Vir Cepheids. The
period-luminosity relation for the Type II Cepheids yielded the
luminosity $\log L/L_{\sun}\sim2.95$. From a low-resolution
spectrum, obtained at maximum brightness, the following
atmospheric parameters were determined: $T_\text{eff}\sim7400$\,K
and $\log g\sim1.38$. This spectrum contains the emission lines
H$\alpha$, \ion{Ba}{II} $\lambda$6496.9, \ion{He}{i}
$\lambda$10830 and Pa$\beta$. Spectral monitoring performed in
2019--2021 showed a significant change in the H$\alpha$ profile
and appearance of CH and CN molecular bands with pulsation phase.
The metal lines are weak. Unlike typical W~Vir variables, the star
shows a strong excess of infrared radiation associated with the
presence of a heavy dust envelope around the star. We modelled the
SED using our photometry and archival data from different
catalogues and determined the parameters of the circumstellar dust
envelope. We conclude that IRAS~02143+5852 is a low-luminosity
analogue of dusty RV~Tau stars.

\end{abstract}


\section{Introduction}\label{intro}

IRAS~02143+5852 ($02^{\mathrm{h}}17^{\mathrm{m}}57\fs8,
+59\degr05\arcmin52\arcsec$, 2000) is an infrared (IR) source
mentioned for the first time by \citet{manch89} as an object with
the far-IR colours similar to those of planetary nebulae. These
authors suggested that the object is in the post-AGB stage of
evolution. From this point on, the star began to be studied
together with other post-AGB candidates. \citet{Omont1993}
considered it as a carbon-rich proto-planetary nebula (PPN).
Observations in the near-IR were carried out by \citet{gl90},
\citet{Fujii02}, \citet{Ueta03} and \citet{Cooper13}.
\citet{Ueta03} pointed out that the $H$ and $K'$ magnitudes showed
a slight variation compared to previous observations. Extended
dust shells have been found in a number of post-AGB stars, however
IRAS~02143+5852 is unresolved at 11\,$\micron$ \citep{meixner99}.
\citet{Gled05} did not detect polarization within errors in the
$J$-band. No H$_{2}$O maser emission was detected by
\citet{suarez07}.

\citet{Fujii02} were the first to obtain optical magnitudes for
the star: $B=14.96$\,mag and $V=13.74$\,mag. The authors assumed
the star to be of F5Ib spectral type with corresponding
temperature $T_{\text{eff}}=6900$\,K, intrinsic colour
$(B\!-\!V)_0=0.33$\,mag and colour excess $E(B\!-\!V)=0.89$\,mag.
\citet{Fujii02} analysed the SED spanning the optical to far-IR
wavelengths and obtained an estimate for the dust shell
temperature $T_{\text{dust}}=205$\,K.

The spectral type of the object is not clear. \citet{KH05}
classified the IRAS~02143+5852 spectrum as an Ae one. Based on a
low-resolution spectrum, \citet{suarez06} stated an F7Ie spectral
type. Analysing the spectrum presented in the appendix of the
above paper, \citet{molina18} found $T_{\text{eff}}=7967\pm91$\,K.
\citet{SP2004} noted that the optical spectrum showed strong
Balmer lines in absorption and no helium lines.

The light variability of the star was discovered based on the All
Sky Automated Survey for SuperNovae (ASAS-SN) data \citep{shap14,
koch17}. In the ASAS-SN catalogue of variable stars
\citep{Jayasinghe19}, it has the designation
ASASSN-V~J021757.82+590552.0, a period of 50.18\,d and an YSO
(Young Stellar Object) variability type. The star is also
contained in the ZTF catalogue of periodic variable stars
\citep{Chen20}, where it is designated as ZTFJ021757.80+590552.1,
has period 25.1385074\,d, and is classified as CepII (Cepheid
variable).

In this paper, we present new photometric and spectroscopic
observations of IRAS 02143+5852 obtained in 2017--2021. These data
allowed us to detect light variability and to study its character,
as well as to determine the parameters ($T_{\textmd{eff}}$, $\log
g$) of the star and to investigate spectrum change. We have used
all currently available IR photometric data to perform the SED
modelling aiming to probe the properties of the circumstellar
material.

The paper is organized as follows. Photometric and spectroscopic
observations and data reduction procedures are described in
Section~\ref{obs}. In Section~\ref{photom} we analyse photometric
data. Section~\ref{spectra} presents the results of low-resolution
spectroscopy. In Section~\ref{evolution} we discuss the
evolutionary status of the star. Section~\ref{sed} is devoted to
the SED modelling. In Section~\ref{disc} we discuss the obtained
results and the star's similarity to known objects.
Section~\ref{concl} presents our conclusions.

\section{OBSERVATIONS AND DATA REDUCTION}\label{obs}

\subsection{\texorpdfstring{$UBVR_\text{C}I_\text{C}$ photometry}{UBVRcIc}}

The photometric observations of IRAS~02143+5852 were carried out
with the FLI ML3041 CCD (2048$\times$2048 pixels, the pixel size
15\,$\micron$) mounted on the 0.6-m telescope of the Star\'{a}
Lesn\'{a} Observatory of the Astronomical Institute of Slovak
Academy of Science (Sl600) in 2018--2019 and with the Andor iKon-L
BV camera (2048$\times$2048 pixels, the pixel size
13.5\,$\micron$) mounted on the new 0.6-m telescope (RC600)
installed at the Caucasian Mountain Observatory (CMO) of the
Sternberg Astronomical Institute of the Moscow State University
(for more details see \citet{berd2020a}) in 2019--2021. Each
detector was equipped with a set of Johnson-Cousins
$UBVR_{\text{C}}I_{\text{C}}$ filters. Standard data reduction
procedures (debiasing, flat-fielding, darking) and aperture
photometry were performed using the \textsc{maxim dl5} package and
self-developed \textsc{python} scripts. The mean photometric
uncertainty is about 0.005-0.010 mag for the $B$, $V$,
$R_{\text{C}}$, $I_{\text{C}}$ bands and up to 0.05 for the $U$
band.

\begin{figure}
\includegraphics[width=\columnwidth]{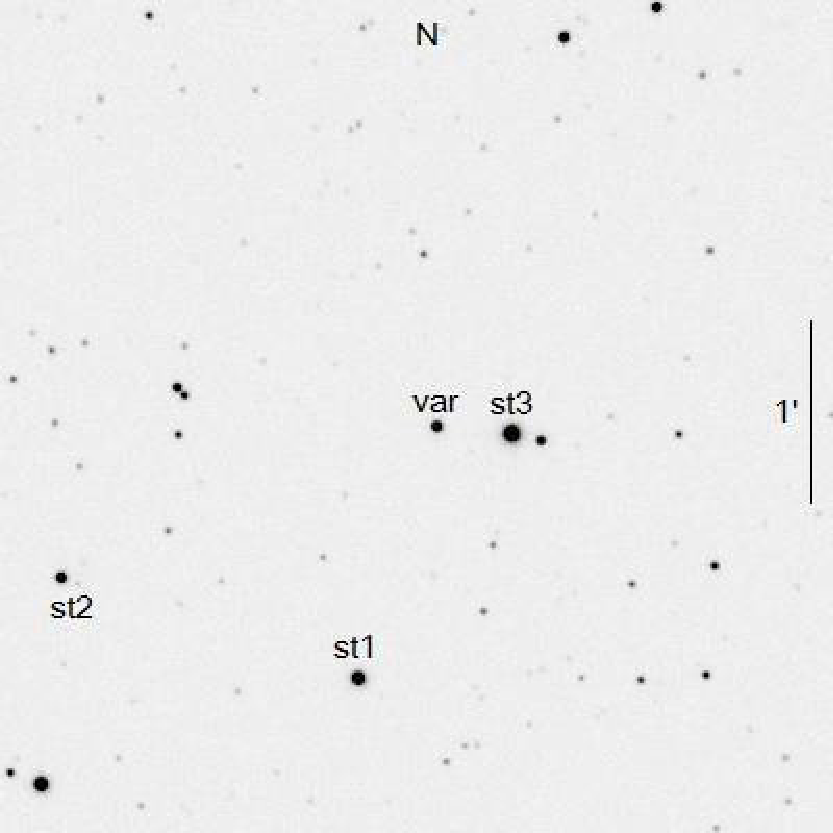}
\caption{The $R_{\text{C}}$-band image of the IRAS~02143+5852 (var) field with
the comparison stars marked. This CCD-frame was obtained with RC600 on February 15, 2020.}
\label{Rc}
\end{figure}

Information on the comparison stars, used for differential
photometry, is presented in Table~\ref{stars}. The magnitudes for
the comparison stars were acquired via referencing to photometric
standards in the S23-246 and S23-436 fields \citep{land13}. The
$R_{\text{C}}$ image of IRAS~02143+5852 and comparison stars is
shown in Fig.~\ref{Rc}. The $UBVR_{C}I_{C}$-photometry is
presented in Table~\ref{phot} where for every night we list the
mean time of observation and magnitudes in each of photometric
bands averaged over 2â€“3 frames.


\begin{table}
\caption{The comparison stars for $UBVR_{c}I_{c}$ photometry.}\label{stars}
\begin{center}
\begin{tabular}{cccccccc}
\hline
Star& $U$&$B$&$V$&$R_{\text{C}}$&$I_{\text{C}}$\\
\hline

st1 &14.841 & 13.932 & 12.629 & 11.840 & 11.052\\
st2 &14.802 & 14.417 & 13.512 & 12.998 & 12.524\\

\hline
\end{tabular}
\end{center}
\end{table}

\subsection{\texorpdfstring{$JHK$-photometry}{JHK}}

The $JHK$-photometry was obtained from 2017 November 6 to 2021
February 18 (73 nights in total) with the ASTROnomical
Near-InfraRed CAMera  (ASTRONIRCAM) \citep{Nadjip17} mounted on
the 2.5-m telescope of CMO. We used the dithering mode to obtain
images in the $JHK$ bands of the MKO--NIR system (Mauna Kea
Observatories Near-InfraRed \citep{Simons02}). Each frame was
automatically reduced and calibrated using the pipeline described
in detail in \cite{ANC2023}. The standard reduction procedures
were performed including the correction for non-linearity and bad
pixels, dark subtraction, flat-fielding and background
subtraction. The instrumental magnitudes for the star were derived
using differential aperture-based photometry with respect to 2MASS
stars (Fig.~\ref{Rc}, Table~\ref{starsIR}) which magnitudes were
converted to the MKO-NIR system following the equations from
\citet{leggett06}. The $JHK$-photometry in the MKO-NIR and 2MASS
systems is presented in Table~\ref{nir}.

As we obtained several frames for each filter during each
pointing, uncertainties were calculated as standard deviations.
The average uncertainties are $\Delta J=0.011$~mag, $\Delta
H=0.010$~mag, $\Delta K=0.012$~mag.


\begin{table}
 \caption{The comparison stars for $JHK$ photometry.}\label{starsIR}
\begin{center}
\begin{tabular}{cccccccc}
  \hline

Star&    $J$(2MASS)& $H$(2MASS)& $K$(2MASS)\\
\hline
St1& 9.977& 9.345& 9.168\\
St3& 9.977& 9.776& 9.705\\

\hline
\end{tabular}
\end{center}
\end{table}


\subsection{Spectroscopy}

A low-resolution optical and near-IR spectrum of IRAS~02143+5852
was obtained on the 3-m telescope of the Lick Observatory (USA)
with the Aerospace Corporation's Visible and Near Infrared Imaging
Spectrograph (VNIRIS) \citep{Rudy21} (0.46--2.5\,$\micron$,
$R\!\sim\!700$) on 2018 October 20 (JD 2458412.5). The spectra
were calibrated using the solar-type standard star HIP~9829.

Low-resolution spectra were acquired on the 2.5-m telescope of CMO
via the Transient Double-beam Spectrograph (TDS) equipped with two
Andor Newton 940P cameras using E2V CCD42-10 detectors, and volume
phase holographic gratings (see \cite{potanin20}). A 1$\farcs$0
slit was used. The log of observations is given in
Table~\ref{sp-obs}. The data reduction procedures including dark
and flat-field correction, cosmic ray removal, two-dimensional
wavelength linearization, background subtraction, and relative
flux calibration using spectrophotometric standards listed in
Table~\ref{sp-obs} were performed using \textsc{python} scripts.


\begin{table}
\caption{Spectroscopic observations of IRAS~02143+5852 on SAI CMO}\label{sp-obs}
\begin{center}
\begin{tabular}{ccccc}
\hline
\textnumero & Date & HJD& Exposure& Standard\\
&yyyy-mm-dd & & time, s& stars\\
\hline

1&2020-01-18&2458867.30& 300$\times$1&BD+25$\degr$4655\\
2&2020-01-20&2458869.28& 300$\times$1&BD+25$\degr$4655\\
3&2020-02-26&2458906.25& 900$\times$1&HR1544\\
4&2020-03-03&2458912.21& 600$\times$3&Hiltner 600\\
5&2020-08-30&2459093.51& 600$\times$1&BD+28$\degr$4211\\
6&2020-09-06&2459099.33& 1200$\times$1&BD+28$\degr$4211\\
7&2020-09-12&2459105.49& 1200$\times$1&HR153\\
8&2020-09-23&2459116.46& 600$\times$1&HR153\\
9&2020-11-01&2459155.34& 600$\times$3&BD+28$\degr$4211\\
10&2020-12-14&2459198.18& 600$\times$3&HR1544\\
11&2021-02-05&2459251.34& 600$\times$2&Feige 66\\
12&2021-02-11&2459257.20& 600$\times$2&HR718\\
\hline
\end{tabular}
\end{center}
\end{table}


\section{Photometric analysis}\label{photom}

\begin{figure*}
\includegraphics[scale=0.7]{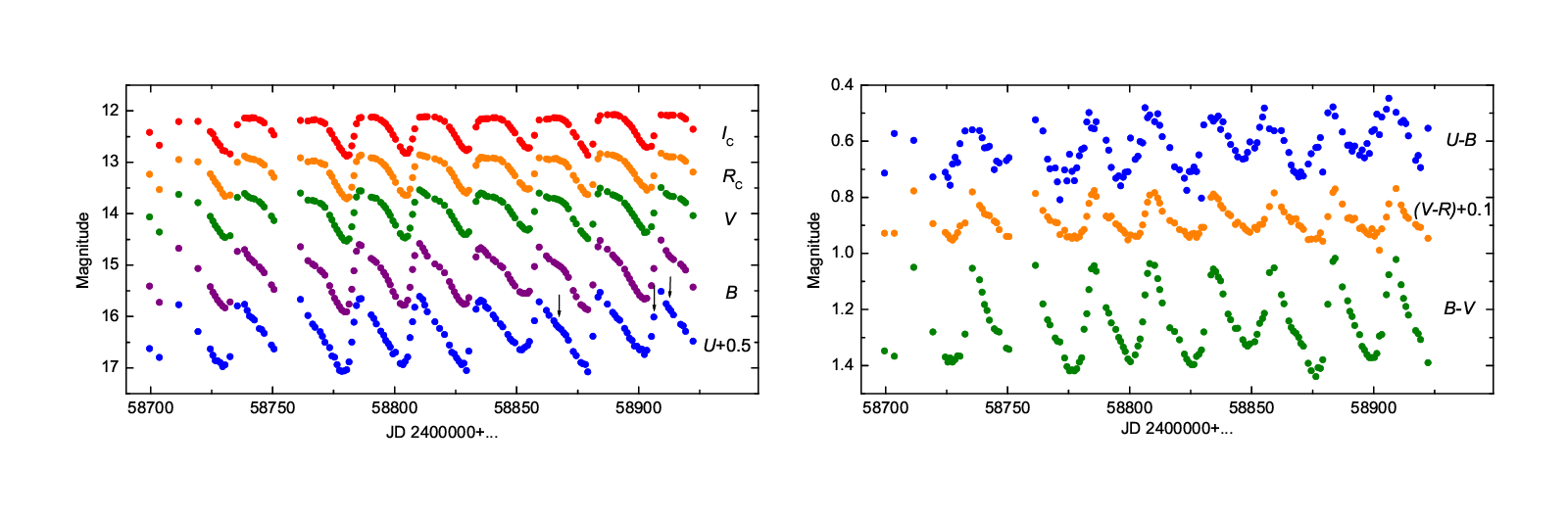}
\caption{The $UBVR_{C}I_{C}$-band light and colour curves covering the period
from August 2019 to March 2020.}\label{lc&ic}
\end{figure*}

In Fig.~\ref{lc&ic} we present the $UBVR_{C}I_{C}$ light and
colour curves for the period from 2019 August to 2020 March which
was best covered by observations. A periodic variation is clearly
seen both in light and colour curves. The shape of light curve
differs from band to band being sawtooth in the $U$ band while the
$R_C$ and $I_C$ light rapidly rises to a flat and long-lasting
maximum. There is a bump on the declining branch of the $B$ and
$V$ light curves. One can also see that deeper minima alternate
with shallower ones and this effect is most pronounced in the $U$
and $B$ bands and in the $B\,-\,V$ colour. The $BVR_{C}I_{C}$
light curves of IRAS~02143+5852 are very similar to those of W~Vir
which is the Type II Cepheid prototype \citep{th07}.

\begin{figure}
\includegraphics[width=\columnwidth]{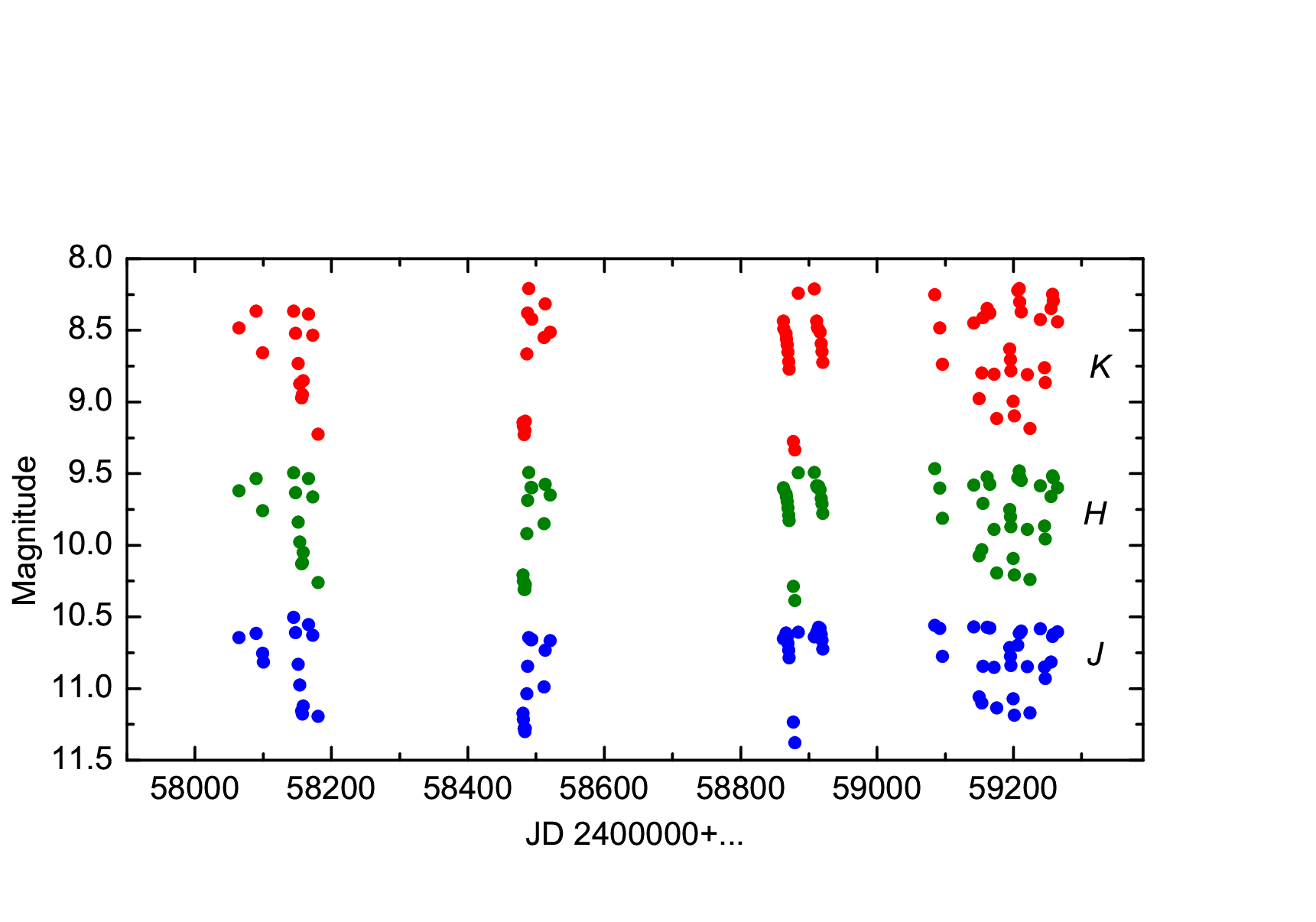}
\caption{The $JHK$-band light curves covering the period from 2017 November to 2021
         February.}\label{LC-IR}
\end{figure}

The near-IR light curves spanning the whole observational period
are shown in Fig.~\ref{LC-IR}. As can be seen, the
$JHK$-brightness varies with a peak-to-peak amplitude of about
1\,mag.

\subsection{Periodicity analysis}

In order to study the periodicities, we have used the
\textsc{winefk} code developed by
V.P.~Goranskii\footnote{\url{http://vgoranskij.net/software/WinEFrusInstruction.pdf}}.
This code implements the well-known method of minimizing the phase
dispersion \citep{LK1965} and the discrete Fourier transform for
arbitrarily distributed time series \citep{Deeming75}.

\begin{figure}
\includegraphics[width=\columnwidth]{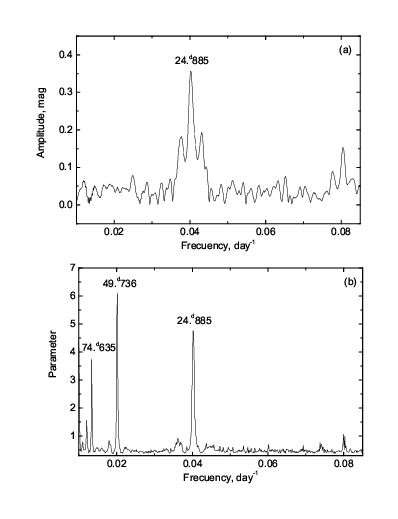}
\caption{Frequency spectra for $V$ data  obtained by the Deeming
(a) and Lafler-Kinman (b) methods. n the bottom panel,
Parameter\,=\,$1/\theta$ is plotted along the ordinate axis, where
$\theta$ is the normalised sum of the squares of the deviations of
each subsequent point from the previous point in the light curve
with a trial period.} \label{fig:period}
\end{figure}

The Fourier spectra of our $V$-band data presented in
Table~\ref{phot} is shown in Fig~\ref{fig:period}. In the
frequency spectrum obtained by the Deeming method a peak
corresponding to a period of 24.885\,d is dominant
(Fig.~\ref{fig:period} (a)). There are two weaker peaks on both
sides of the primary one which turn out its one-year aliases.
Taking into consideration all our $UBVR_CI_C$ data for the
2018--2021 interval we derived $P=24.885\pm0.150$\,d.


\begin{figure*}
\includegraphics[scale=0.7]{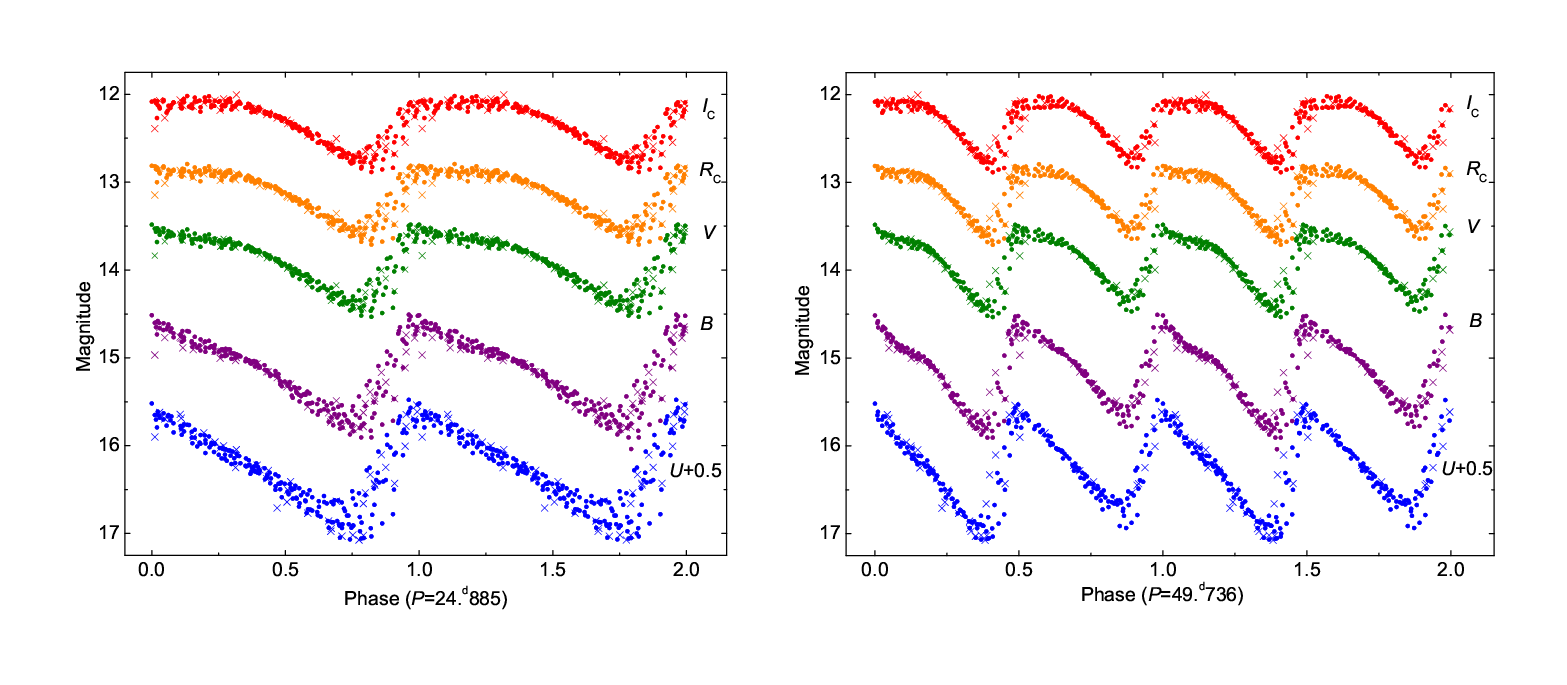}
\caption{The $UBVR_{C}I_{C}$ light curves folded on the periods
of 24.885~d (left panel) and 49.736\,d (right panel). Filled circles
represent the CMO data, crosses -- the data obtained in Slovakia.}
\label{fig:plc}
\end{figure*}


Analysing the times of maxima allowed us to revise the ephemeris to be:
\begin{equation}
\begin{array}{c}
\text{JD(max)}=2458660.352+24.885\times E
\end{array}
\end{equation}
where 'maximum' is defined as a peak at the end of the rising
branch.

The Lafler-Kinman method gives a frequency spectrum that contains
peaks corresponding to the period of 24.885\,d, and its multiples
of 49.736 and 74.635\,d (Fig~\ref{fig:period} (b)).

The light curves folded on the 24.885\,d period show a significant
scatter ($>$0.1\,mag) at phase 0.6--1.0 which is much larger than
observation errors indicating that there are cycle-to-cycle
variations. Folding on twice the above period reveals some details
in light curves, particularly, the alteration of deep and shallow
minima. The $UBVR_{\text{C}}I_{\text{C}}$ light curves folded on
periods of 24.885 and 49.736\,d are shown in Fig.~\ref{fig:plc}.

\begin{figure*}
\begin{center}
\includegraphics[scale=0.4]{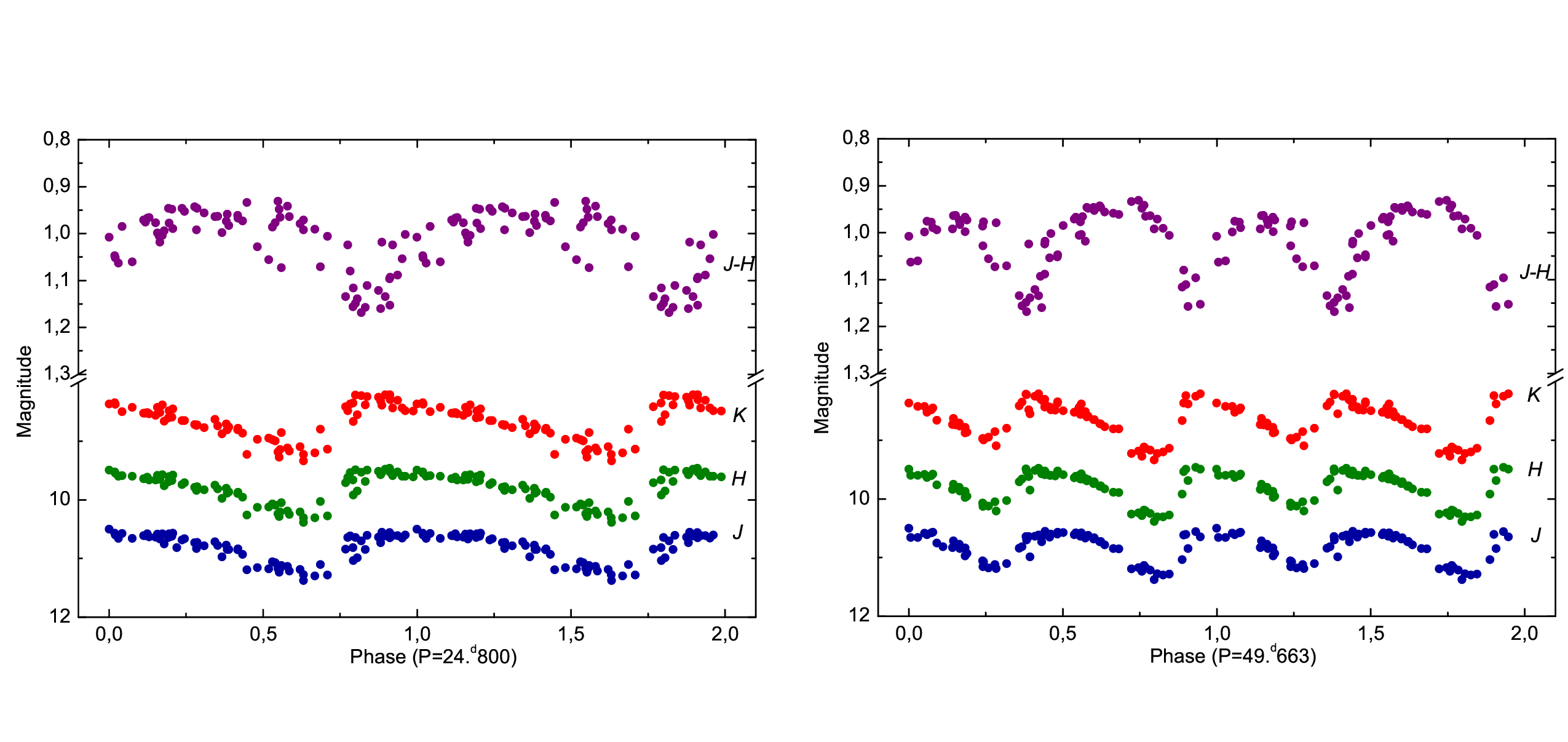}
\caption{The $JHK$ light and $J\!-\!H$ colour curves folded on the periods
of 24.793\,d (left panel) and 49.690\,d (right panel).}\label{phaseIR}
\end{center}
\end{figure*}

The near-IR data are less numerous than the optical data, so, it
is harder to determine reliable period. The power spectrum derived
by the Lafler-Kinman method for periods in the range 10--100\,d is
dominated by the peaks which correspond to the periods of
24.819\,d and 49.690\,d ($J$-band), 24.793\,d and 49.690\,d ($H$
and $K$ bands). The $JHK$ light and $J\,-\,H$ colour curves folded
on the period $P$=24.819\,d and on nearly twice the period
$P$=49.690\,d are shown in Fig.~\ref{phaseIR}.

The $JHK$ phase curves are similar in shape showing a rapid rise
to maximum and gentle slope. The phase curves folded on twice the
period demonstrate the pattern of minima alternating in depth
which is typical for the RV~Tau type variables. The $J\,-\,H$
colour reaches maximum at maximum light, so, the star is redder
when brighter.


A brief summary of characteristics yielded from the photometric
study, namely, the periods derived for different photometric bands
as well as maximum brightness, peak-to-peak amplitudes, and the
number of observations ($N$), is given in Table~\ref{res}.


\begin{table}
\caption{Summary of the photometric study of IRAS~02143+5852 in 2018--2021.}\label{res}
\begin{center}
\begin{tabular}{ccccc}
\hline
Band&Brightness of&Peak-to-peak &$P$& $N$\\
&maximum&amplitude\\
&(mag)&(mag)&(days)&\\
\hline

$U$& 15.09& 1.54&24.885& 251\\
$B$& 14.64& 1.17&24.885& 251\\
$V$& 13.55& 0.91&24.885& 251\\
$R_{\text{C}}$& 12.87&0.76&24.885& 251\\
$I_{\text{C}}$& 12.12&0.74&24.885& 251\\
$J$(MKO-NIR)& 10.54& 0.75&24.819&73\\
$H$(MKO-NIR)& 9.49& 0.82&24.793&73\\
$K$(MKO-NIR)& 8.21& 0.99&24.793&73\\

\hline
\end{tabular}
\end{center}
\end{table}


\begin{figure}

\begin{center}

\includegraphics[scale=0.7]{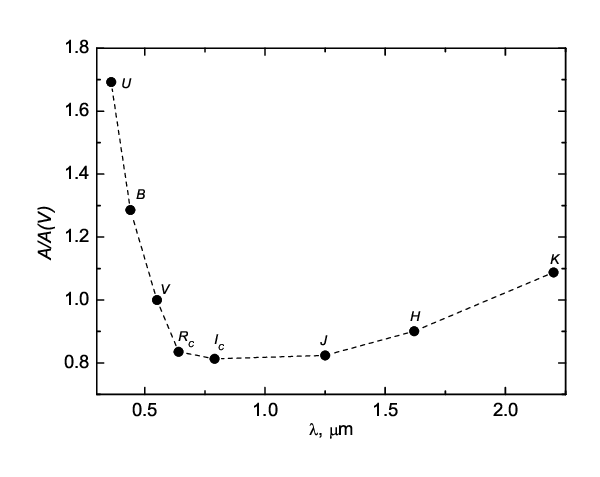}
\caption{The ratio of the amplitude in $UBVR_{C}I_{C}JHK$-bands
to the amplitude in the $V$-band.}\label{ampl}

\end{center}

\end{figure}


The data presented in Table~\ref{res} and Fig.~\ref{ampl}
indicates that the  oscillation amplitude is the largest in the
$U$-band, decreases with increasing wavelength in the optical
range up to the $I_C$-band, and then begins to increase in the
near-IR range from $J$ to $K$. The ratio of $A(K)$ to $A(V)$ is
1.08. This is an unexpected result, since in pulsating variables
the amplitudes of oscillations in the near-IR range are usually
lower than in the optical. For example, in classical Cepheids with
periods $P>20$\,d, the ratios of the amplitudes $A(J)$, $A(H)$ and
$A(K)$ to $A(V)$ are approximately 0.4 \citep{Inno15}. As we show
below, IRAS~02143+5852 exhibits a significant excess of near-IR
emission. As almost half of the $H$-band emission arises from dust
and pulsations do affect the emission from the dust envelope, it
seems credible to relate the observed
near-IR-amplitude--wavelength dependence to the presence of dust.

\subsection{Colour-colour diagrams}\label{col-col}

\begin{figure}
\begin{center}
\includegraphics[width=\columnwidth]{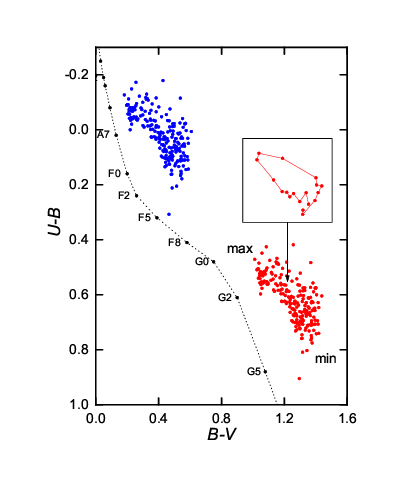}
\caption{{\it U-B, B-V} colour-colour diagram. Red and blue dots are observed
and dereddened with $E(B\!-\!V)$=0.83\,mag data, respectively. The supergiant
theoretical sequence is plotted with black dots connected by dotted line.
In the inset, the data for one pulsation cycle (JD2458870.2--2458896.2) are shown.}\label{twocol}
\end{center}
\end{figure}

The location of IRAS~02143+5852 in the $U\!-\!B, B\!-\!V$
colour-colour diagram is shown in Fig.~\ref{twocol} along with the
sequence of supergiants from \citet{strai82}. The observed colours
vary from the reddest in minimum light to the bluest in maxima
which is common to temperature oscillations related to pulsations.
Analysing spectroscopic data (see Section~\ref{spectra}), we found
that at light maximum the star has a spectral class of
approximately F0I, for which the normal colour index is
$(B-V)_{\text{0}}=0.20$\,mag \citep{strai82}. The observed colour
at light maximum is $B-V=1.03\pm0.02$\,mag, and therefore the
colour excess can be estimated as $E(B-V)=0.83\pm0.02$\,mag. In
the colour-colour diagram (Fig.~\ref{twocol}), the dereddened
colours are located above the sequence of supergiants, indicating
that the star has some ultraviolet excess, which is typical of
W~Vir stars and may be due to low metallicity \citep{strai82}. As
we discuss below, IRAS~02143+5852 also has low metal content.


\begin{figure}
\begin{center}
\includegraphics[width=\columnwidth]{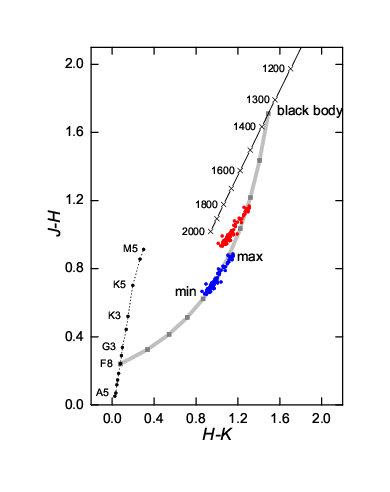}
\caption{$J$-$H, H$-$K$ colour-colour diagram. Red and blue dots
are observed and dereddened with $E(B\!-\!V)=0.83$\,mag data,
respectively. The solid black line shows the loci of a blackbody
with temperature ranging from 1200 to 2000\,K. The dotted black
line shows the supergiant sequence. The grey squares connected by
a thick grey line correspond to a combination of an F8I star and a
1350\,K blackbody, where each square is calculated with a step of
0.1 in terms of the fractions of contributions from individual
components to the total emission in the $H$-band. }\label{fig:nir}
\end{center}
\end{figure}


The star demonstrates a significant near-IR excess. In
Fig.~\ref{fig:nir} we plot the $J\!-\!H, H\!-\!K$ colour-colour
diagram for IRAS~02143+5852, the supergiant sequence from A5 to M5
and a blackbody with temperature in the range from 1200 to
2000\,K. The $J-H$ and $H-K$ colours for supergiants were taken
from \citet{koorn83} and converted to the 2MASS system using
equations from \citet{Carpenter01} and then to the MKO-NIR system
using equations from \citet{leggett06}. The star's colours
dereddened with $E(B\!-\!V)=0.83$\,mag occupy the location which
can be attributed to a sum of radiation from a $\sim$F5 supergiant
and hot dust with $T_{\text{dust}}\sim1300$\,K. Note that the dust
contribution is bigger in maximum light than in minimum.

The star also has a considerable excess of far-IR radiation
related to cold dust. We present the circumstellar dust shell
modelling in Section~\ref{sed}.

\section{Spectral features and stellar parameters}
\label{spectra}

\begin{figure}
\begin{center}
\includegraphics[width=\columnwidth]{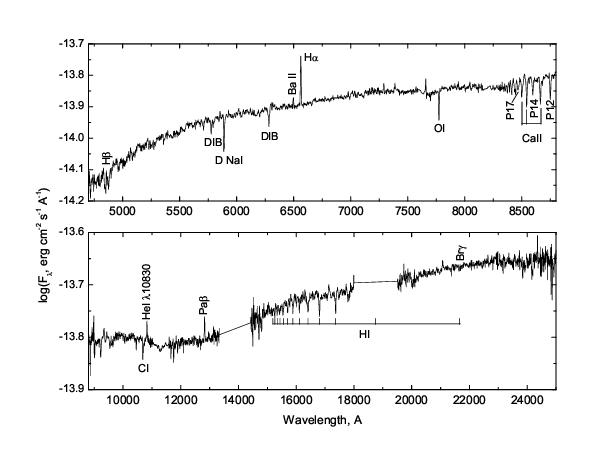}
\caption{The flux-calibrated low-resolution spectrum taken at the
Lick Observatory covering 4700--25000\,\AA. The main spectral features are indicated.}\label{sp}
\end{center}
\end{figure}

The flux-calibrated spectrum obtained on 2018 October 20 near the
maximum brightness ($\phi=0.07$) is shown in Fig.~\ref{sp}.
Because of low resolution in the short wavelength region most of
lines are blended, which makes the analysis difficult. The most
prominent details in the spectrum are the emission lines:
H$\alpha$, \ion{Ba}{II} $\lambda$6496.9, \ion{He}{i}
$\lambda$10830 and Pa$\beta$. The H$\beta$ line has also an
emission component.

The main absorption features are D \ion{Na}{I}, diffusion
interstellar bands (DIBs) at $\lambda$5780 and $\lambda$6280, the
\ion{O}{I} triplet at $\lambda\lambda$7771--74, the \ion{Ca}{II}
triplet ($\lambda\lambda$8498, 8542, 8662), the Paschen hydrogen
lines, in particular P12, P14, P17, and the Brackett hydrogen
lines.

The equivalent widths ({\it EWs}) of the most prominent lines are
presented in Table~\ref{ew}. We consider {\it EW} negative for
absorption and positive for emission. The measurement error is
estimated to be about 10 per cent.


\begin{table}
\begin{center} \caption{Equivalent widths of absorption and emission lines
in spectrum obtained on 2018 October 20.} \label{ew}
\begin{tabular}{ccc}
\hline
$\lambda_\text{lab.}$ & Identification & $EW$\\
(\AA) && (\AA)\\
\hline

6496.9&\ion{Ba}{II}&0.53\\
6563&\ion{H}{I}&3.20\\
7771-74&\ion{O}{I}&-1.41\\
8469&P17&-1.18\\
8498&\ion{Ca}{II}&-1.85\\
8542&\ion{Ca}{II}&-2.23\\
8600&P14&-1.55\\
8662&\ion{Ca}{II}&-2.28\\
8753&P12&-1.73\\
10830&\ion{He}{i}&1.33\\
12818&P$\beta$&2.67\\
\hline
\end{tabular}
\end{center}
\end{table}

We have estimated $T_{\text{eff}}$ and $\log g$ of the star, using
empirical relations from  \citet{molina18} and \citet{mant91}.
\citet{molina18} proposed a functional relationship to assess the
value of $\log g$:

\begin{equation}
\begin{array}{c}
\log g=(2.20\pm0.20)-(0.58\pm0.13)(EW({\text{\ion{O}{I}}})),
\end{array}
\end{equation}

where $EW({\text{\ion{O}{I}}})$ is the equivalent width of the
\ion{O}{I} triplet lines at $\lambda\lambda$7771--74. For
IRAS~02143+5852 we have derived $\log g=1.38\pm0.38$ which
corresponds to the Iab luminosity class according to the
calibration of \citet{strai82}. But it is necessary to keep in
mind that the strength of the \ion{O}{I} triplet lines also
depends on other atmospheric parameters \citep{kovt11},
particularly on metallicity, which can be defined reliably only
via the analysis of high-resolution spectra.

As was shown by \citet{mant91}, the ratio {\it Ca/P} where {\it
Ca} is the sum of {\it EWs} of the \ion{Ca}{II} IR triplet lines
and {\it P} is the sum of {\it EWs} of P12, P14, and P17, can be
used to estimate the effective temperature of a star. Just as was
reported by \citet{mant91} for the spectra of RV~Tau stars, the
components of the \ion{Ca}{II} IR triplet are blended with P13,
P15, and P16 in our spectra. Table~\ref{ew} reports the measured
{\it EWs} of P12, P14, P17 and \ion{Ca}{II} triplet lines. In
accordance with eq.~1 from \citet{mant91},

\begin{equation}
\begin{array}{c}
\log T_{\text{eff}}=3.90-0.20 \log(\text{Ca/P})\pm 0.07,
\end{array}
\end{equation}
we derived $T_{\text{eff}}=7400_{-1152}^{+1248}$\,K.

\begin{figure}
\begin{center}
\includegraphics[width=\columnwidth]{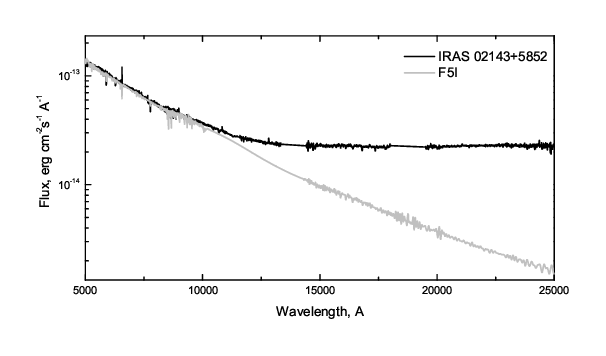}
\caption{The Lick Observatory spectrum dereddened with $E(B\!-\!V)=0.83$\,mag
(black line) and the synthetic spectrum of an F5I star taken from \citet{Pickles98}
(grey line).}\label{sp1}
\end{center}
\end{figure}

The calibration of \citet{flower96} ascribes the intrinsic colour
$(B-V)_0=0.20$\,mag to a supergiant with $T_{\text{eff}}=7400$\,K.
Our spectrum was obtained near maximum light when the mean value
of $B-V$ is about 1.03\,mag, so, the colour excess can be
estimated as 0.83\,mag. The spectrum dereddened with
$E(B-V)=0.83$\,mag is shown in Fig.~\ref{sp1} together with the
synthetic spectrum of an F5I star taken from the stellar spectral
flux library presented by \citet{Pickles98}. As one can see, the
5000--7500\,\AA\ spectrum of IRAS~02143+5852  corresponds well to
an F5I type spectrum, whereas at longer wavelengths there is a
significant excess of radiation compared to a standard star.
According to the calibration of \citet{flower77}, an F5I star has
the temperature $T_{\text{eff}}=7000$\,K which falls within the
range of temperatures inferred from the {\it Ca/P} ratio.


\begin{figure}
\begin{center}
\includegraphics[width=\columnwidth]{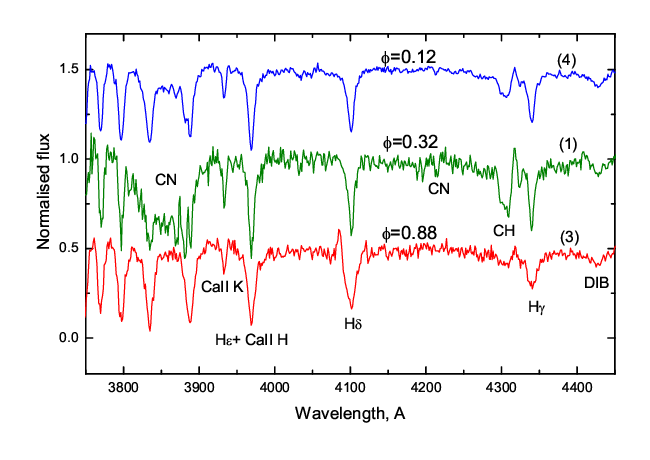}
\caption{The spectrum in the 3750--4450\,\AA\ wavelength range at different
pulsation phase. The numbers in parentheses correspond to the order number of spectra
as it is defined in Table~\ref{sp-obs}.}\label{fig:bluesp}

\end{center}

\end{figure}

We were interested to study spectral changes depending on the
pulsation phase. So, we obtained 12 spectra in the
3500--7500\,\AA\ range at different pulsation phases on the 2.5-m
telescope of CMO with TDS (Table~\ref{sp-obs}). The main
characteristics of the IRAS~02143+5852 spectrum are the following:

\begin{itemize}

\item The optical spectrum is dominated by the Balmer lines.
H$\alpha$ and H$\beta$ have emission components which vary with
pulsation phase. The H$\alpha$ emission appears near minimum
light, persists on the rising branch until maximum and then
weakens rapidly on the declining branch and becomes double-peaked
with a peak-to-peak separation of about 3.6\,\AA\ or 164\,\kms.

Strong hydrogen emission lines observed during rising light are
characteristic of population II Cepheids and their appearance is
explained by the propagation of a shock wave through the expanding
outer layers of stellar atmosphere \citep{Abt54, Whitney56,
Wallerstein59}.

\item In the blue part of the spectrum there are molecular bands
of CH ($\lambda$4261, $\lambda$4280, $\lambda$44323) and CN
($\lambda$4197, $\lambda$4214, $\lambda$4216). The strength of the
bands varies with pulsation phase and is the largest near the
middle of the declining branch. Fig.~\ref{fig:bluesp} shows the
blue spectral region with the molecular bands mentioned above.

\end{itemize}


\begin{figure}
\begin{center}
\includegraphics[width=\columnwidth]{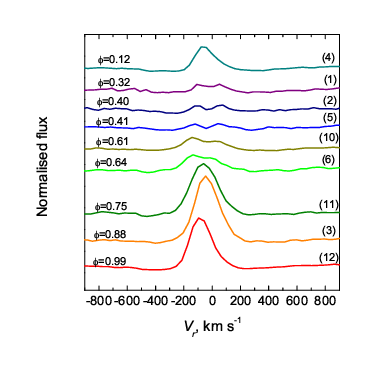}
\caption{The evolution of the H$\alpha$ profile with pulsation phase. The numbers
in parentheses indicate the order number of spectra as it is defined in Table~\ref{sp-obs}.}\label{fig:Ha}
\end{center}
\end{figure}


IRAS~02143+5852 is found to have weaker metal lines in its
spectrum compared to stars of normal metallicity. In order to
illustrate this fact, we show the spectra of IRAS~02143+5852
obtained near minimum and maximum light as well as the spectra of
HD~17971 (F5Ib) and SAO~37370 (F0Ib) obtained with TDS, so, all of
them have the same spectral resolution.


\begin{figure*}
\begin{center}
\includegraphics[scale=1.0]{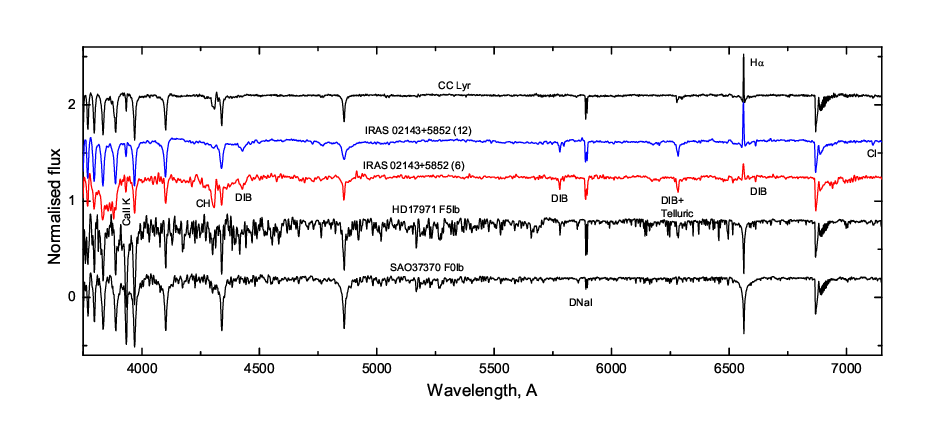}
\caption{The spectra of IRAS~02143+5852 at two phases of pulsation cycle --
near the minimum (6) and near the maximum (12) light -- and the spectra of
HD~17971 (F5Ib), SAO~37370 (F0Ib) and CC Lyr. The spectra are normalised
to continuum and arbitrarily shifted along the vertical axis.}\label{dvd}
\end{center}
\end{figure*}

We have also compared the spectrum of IRAS 02143+5852 with that of
the star CC~Lyr, which is a Type II Cepheid \citep{Harris85}. It
has the pulsation period $P=24.01$\,d \citep{berd2020b} which is
close to that of IRAS~02143+5852. In Fig.~\ref{dvd} we present the
spectrum of CC~Lyr which we obtained on 2020 March 8 when the star
was on the ascending branch of light curve ($\phi=0.76$). The
spectrum has much in common with that of IRAS~02143+5852: the
appearance of the Balmer series is nearly the same, the metal
lines are also weak. Like IRAS~02143+5852, CC~Lyr displays strong
H$\alpha$ emission and the CH band at $\lambda$4280.

\cite{hw84} noted that 'because of their peculiarities, the
spectra of Type II Cepheids are difficult to classify by direct
comparison with MK standards'. Based on HI lines \citet{hw84}
ascribed an F4-F8 spectral type to CC~Lyr depending on the
pulsation phase; from metal lines they deduced an A type and also
pointed to the presence of the CH band. So, according to
\citet{hw84} the spectral classification of CC~Lyr at maximum
light is hF4mA:CH+1. We can state that the spectra of
IRAS~02143+5852 and CC~Lyr are similar in their peculiarity.

\section{Luminosity, distance and evolutionary status}\label{evolution}

IRAS~02143+5852 shows many important similarities to long-period
Type II Cepheids, e.g., the shape of light curves, the pulsation
period, the emergence of Balmer emission, among other observed
photometric and spectroscopic properties. We therefore decided to
estimate the star's luminosity from the period-luminosity relation
established for this type of pulsating variables. We applied the
relation $M_{V}=-0.61-2.95\log P+5.49(V-R)_0$ proposed by
\citet{Alcock98} for the long-period ($0.9 < \log P < 1.75$) Type
II Cepheids.

The mean observed colour of IRAS~02143+5852 is
$B-V=1.25\pm0.11$\,mag. After correcting for reddening with
$E(B-V)=0.83$\,mag we have $(B-V)_0=0.42$\,mag which corresponds
to $(V-R)_0=0.39$\,mag according to the calibration of
\citet{strai92}. This value of $(V-R)_0$ leads to
$M_{\text{V}}=-2.64$\,mag. Accounting for bolometric correction
$BC=-0.007$\,mag \citep{flower96} we derive
$M_{\text{bol}}=-2.65$\,mag and $\log L/L_{\sun}=2.95$.

As IRAS~02143+5852 is positioned as a post-AGB star, let us
compare the parameters of the star with the predictions inferred
from the evolution models for post-AGB objects presented by
\citet{bert16}. Typical post-AGB objects with initial masses on
the Zero Age Main Sequence (ZAMS) between 0.8 and 8.0 $M_{\sun}$
at the stage after shedding their shells on AGB have final masses
of 0.52-0.85 $M_{\sun}$ and luminosities $\log L/L_{\sun}$ ranging
from 3.4 to 4.2, respectively, which is significantly higher than
that obtained for IRAS~02143+5852.

Taking into account the new data presented here and our
presumption of IRAS~02143+5852 being a W~Vir star, we have turned
to the evolutionary models of W~Vir pulsating variables. The
evolutionary status of these stars was discussed more than once,
e.g., by \citet{Gingold74, Gingold76}. \citet{Gingold85} and
\citet{Fadeyev20} summarized the previous work done in this area.
On the basis of self-consistent stellar evolution models and
nonlinear stellar pulsation calculations, \citet{Fadeyev20}
concluded that W~Vir pulsating variables are the low-mass post-AGB
stars that experience the final helium flash. He found that a set
of evolutionary models with masses $M = 0.536M_{\sun}$,
$0.530M_{\sun}$ and $0.526M_{\sun}$ that experience the loop in
the Hertzsprung-Russel diagram, due to the final helium flash, can
provide a solution of hydrodynamic equations which describes
radial oscillations of W~Vir stars. The $\log L/L_{\sun}=2.95$
found for IRAS~02143+5852 falls into the range of luminosities
predicted for the above masses.

The results of the Gaia mission for IRAS~02143+5852 are confusing.
The parallax of the star listed in the Gaia DR2 turned out
negative ($\pi=-0.8487\pm 0.4583$ mas; \citet{gaia18}). The
distance based on these data is $d=3606^{+2032}_{-1331}$\,pc
\citep{bj18}. The parallax $\pi=1.3637\pm 0.2892$ mas introduced
in the Gaia DR3 \citep{gaia22} brings the star much closer to us.
The distance estimate for this parallax is
$d=854^{+276}_{-175}$\,pc \citep{bj21}. At this distance, the star
would have had a luminosity of about 16$L_{\sun}$, however, which
contradicts its evolutionary status.

\section{The model of the circumstellar dust envelope and modelling aspects}\label{sed}

The SED of the object demonstrates a prominent IR excess. Thus,
assuming that the excess is caused by the radiation from heated
dust, we estimated the circumstellar dust envelope parameters via
the simulations with \textsc{radmc-3d} \citep{Dullemond12}.

Our modelling was based on the following sets of observational data: \par
\begin{itemize}
    \item $UBVR_\mathrm{c}I_\mathrm{c}JHK$ photometry. We used the absolute
    flux calibrations from \citet{strai92} for the $U$, $B$, $V$ bands,
    \citet{Bessell79} for the $R_\mathrm{c}$, $I_\mathrm{c}$ bands and \citet{TV05}
    for the $J$, $H$, $K$ bands.
    \item The observed fluxes at 65 and 90\,$\micron$ from the AKARI/FIS Bright
    Source Catalogue. The data at 140 and 160\,$\micron$ were omitted because
    of their low quality.
    \item The observations at 9 and 18\,$\micron$ from the AKARI/IRC Point
    Source Catalogue.
    \item The $W1$, $W2$, $W3$, and $W4$ magnitudes from the AllWISE
    Data Release \citet{cutri13}. The fluxes were calibrated according
    to \citet{Jarrett11}.
    \item The flux densities in the $A$, $C$, $D$, and $E$ bands of
    the {\it Midcourse Space Experiment} ({\it MSXC6}).
    \item The average non-colour corrected flux densities at 12, 25
    and 60\,$\micron$ from the IRAS Point Source Catalog v2.1 (PSC).
    The data at 100\,$\micron$ were excluded as only the upper limit is presented.
\end{itemize}

\begin{figure}
\begin{center}

\includegraphics[width=\columnwidth]{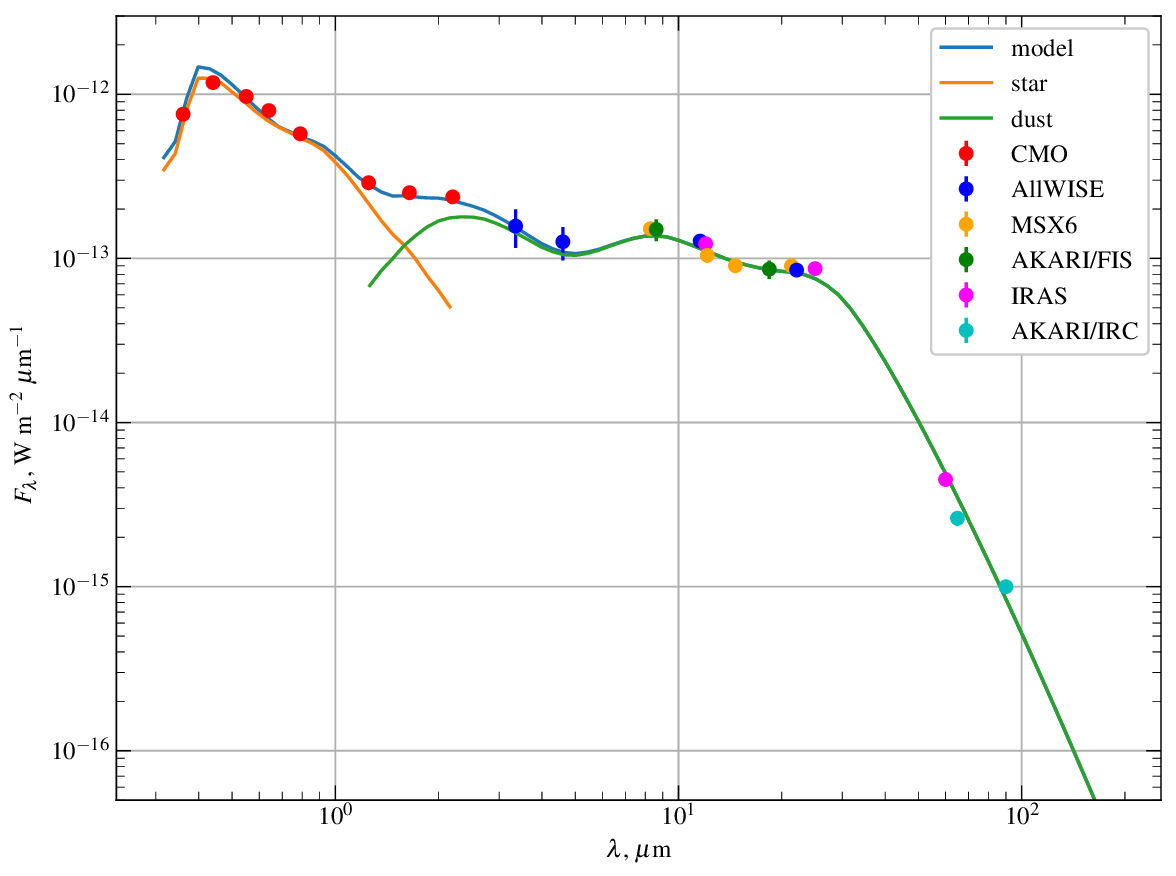}
\caption The SED of IRAS~02143+5852. The blue curve represents the
resulting fit corresponding to the model described in the text,
the orange one -- the SED of the central star reprocessed by the
dust, the green one is the radiation from the circumstellar dust
shell. The symbols depict the observational data points. For some
data the error bars are smaller than the size of the symbols.

\label{fig:SED}

\end{center}
\end{figure}

We modelled the SED of the star using our
$UBVR_{\text{C}}I_{\text{C}}JHK$ data for maximum light and the
data available from different catalogues listed in
Table~\ref{tab:sed}. We used the initial model atmosphere
parameters $T_{\text{eff}}=7460$\,K and $\log g$=1.38 found in
this study which correspond to an F5I star from \citet{Pickles98}.
An example of the fitting SED derived for IRAS~02143+5852 is shown
in Fig.~\ref{fig:SED}.


\begin{table*}
 \caption{Photometric data used to construct the SED for IRAS 02143+5852.}
 {\small
 \label{tab:sed}
 \begin{tabular}{ccccc}
  \hline
Telescope&  Wavelength ($\micron$)& Flux (Jy) & Obs. time & Referense\\

  \hline

WISE     &   3.4, 4.6, 12, 22&  0.586,
0.836, 5.685, 13.803
 &2010&\citet{cutri13}\\
AKARI PSC&   9, 18 & 3.712, 9.705 &2006-2007&\citet{mur07}\\
AKARI BSC&  90, 140, 160 &2.697, 1.297, 0.361 &2006--2007  &\citet{mur07}\\
IRAS&   12, 25, 60 &5.9, 18.1, 5.39    &1983&\citet{hw88}\\
MSX& 8.28, 12.13, 14.65, 21.34&3.67, 5.117, 6.044,
15.62&1996&\citet{egan03}\\

 \hline
\end{tabular}
}
\end{table*}


The object is situated close to the galactic plane with the
galactic latitude of about $-2^{\circ}$. According to \citet{bj18}
the distance to the object is within the confidence interval of
about 2272--5636\,pc. These two facts lead to large interstellar
extinction which distorts the spectral energy distribution of the
object.

A distance estimation error causes an uncertainty in the SED
corrected for interstellar extinction. We decided to use the
period-luminosity relation for W~Vir variables \citep{Alcock98} to
estimate the distance to the object. Our approach to this problem
allowed us to reconcile the luminosity derived from this relation,
distance to the star and the luminosity calculated by integrating
the SED corrected for interstellar extinction.

This method is as follows. The period-luminosity relation, when
averaged over the period, gives a bolometric magnitude of
$-2.65$\,mag, corresponding to a luminosity of $910 \;L_{\sun}$.
Comparing the bolometric magnitude to the actual observations
yields a distance of 2470\,pc.

To take into account interstellar extinction we used dust maps by
\citet{dustmaps} and the interstellar extinction law from
\citet{Cardelli89} and \citet{Donnell94} assuming $R_V$ = 3.1 as a
mean for the Galaxy.

To estimate the distance we applied the photometry averaged over
the period and assumed that the fluxes corresponding to the {\it
AKARI}, {\it WISE}, {\it MSX} and {\it IRAS} data did not change
essentially during the pulsation cycle. This assumption needs to
be explained. It was shown by \citet{VCrB} for a Mira-type star
V~CrB that the main parameters of the model dust shell ($\tau$,
$R_{in}$, $R_{out}$ and dust properties) do not depend on whether
the maximum or minimum light SED is fitted. V~CrB varies with an
amplitude of $>4$\,mag in $B$ and $\sim 0.8$\,mag in $KLM$ and a
period of 355.2\,d, so, even if the case of such large-scale
variations does not require different dust shell models for
minimum and maximum light, we can assume the constancy of IR
fluxes during the pulsation cycle for IRAS~02143+5852.

The temperature of the internal star is of great importance for
simulating. Since we have the spectrum at maximum light obtained
at the Lick Observatory we were able to evaluate the corresponding
temperature. The value turned out to be roughly 7400 K. Thus, the
SED modelling was carried out for the phase of the maximum
brightness. This means that unlike the case of determining the
distance when we considered the average SED, for the modelling we
applied the photometry obtained at maximum light. We corrected
these data for interstellar extinction using the distance
estimated as was described above. As mentioned earlier, we presume
that the fluxes corresponding to the {\it AKARI}, {\it WISE}, {\it
MSX} and {\it IRAS} data do not change significantly during the
pulsation cycle. So, we let ourselves adopt these observational
data not only for the average brightness but also for the maximum
light.

The SED of IRAS~02143+5852 has a non-trivial profile, such that
the flat SED in the IR domain can not be explained by the presence
of only one dusty spherical shell with a simple dust density
distribution $\rho \propto r^{-\alpha}$. A series of simulations
showed that in order to reproduce the observed SED we should
consider at least three nested spherical layers that correspond to
the different stages of mass loss. The idea of multiple shells has
been supported by observational evidence for AGB stars (see, for
example, the studies on CW~Leo by \citet{Mauron00} or
\citet{Cernicharo15}).

Thereby, following the law of parsimony, we considered three
shells (three components of the composite dust envelope) in
further modelling, which was carried out under the following
assumptions:

\begin{itemize}
    \item The SED of the star is assumed to be that of an F5I star.
    \item The components of the circumstellar dust envelope are spherical layers
    located symmetrically about the star.
    \item $\alpha=2$ for each spherical shell, since this describes the simplest
    case of the constant mass-loss rate and expansion velocity.
    \item Dust grains are spherical. Hence, the opacity coefficients were
    calculated in accordance with the Mie theory.
    \item Since there is no indication of the chemical composition, we used
    amorphous carbon dust grains to avoid additional ambiguity. The corresponding
    optical constants were taken from \citet{Suh00}.
\end{itemize}

Fig.~\ref{fig:SED} represents the best-fitting model SED with
minimal normalized deviations from the observed data.


The parameters of the corresponding model dust shells, namely, the
distance from the stellar center to the inner and outer
boundaries, the radius of dust grains, the optical depth, the dust
mass and the mass-loss rate, are listed in Table~\ref{tab:params}.

\begin{figure}
\centering
\includegraphics[width=\columnwidth]{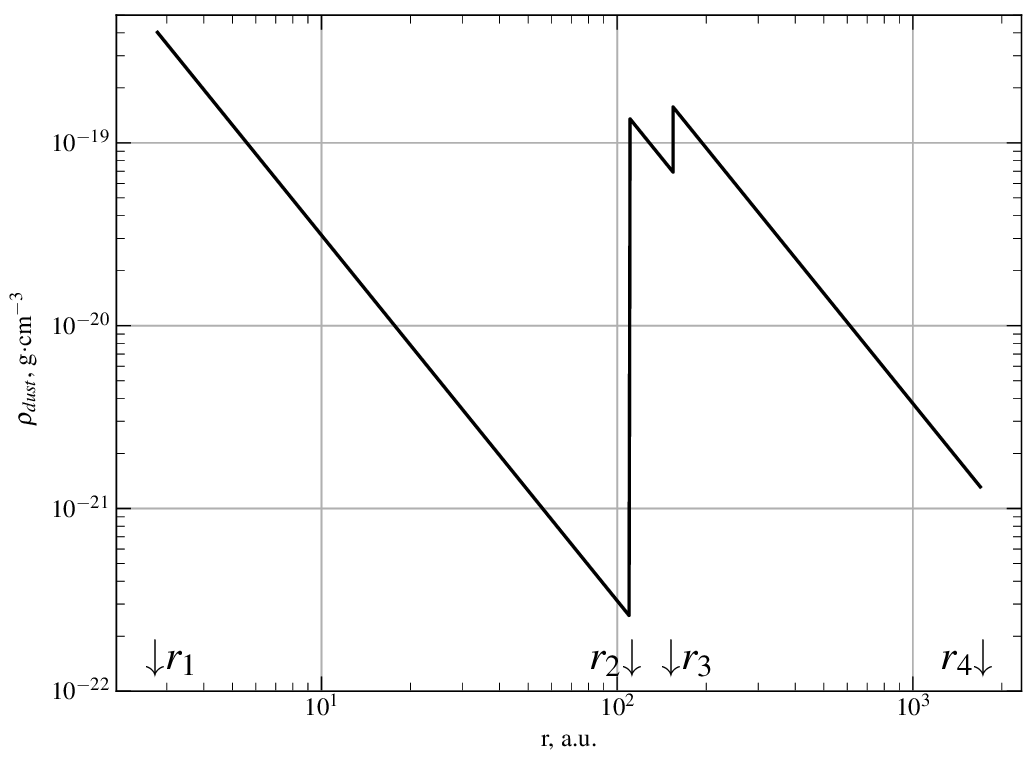}
\caption{The model dust density distribution in the envelope
plotted on logarithmic scale.  The values  $r_1$, $r_2$, $r_3$ and
$r_4$ -- the boundary positions of the dust envelope components.
$r_1$ = 2.5\,au, $r_2$ = 110\,au, $r_3$ = 146\,au and $r_4$ =
1750\,au according to Table~\ref{tab:params} where the model
parameters are listed.} \label{fig:rhod}
\end{figure}

Our estimates of the mass-loss rate were derived under the
assumption that the expansion velocity and the gas-to-dust ratio
are equal by the order of magnitude to the typical values found
for the circumstellar envelopes of AGB stars:
$v_\mathrm{e}$=10\,\kms\ and
$\rho_{\mathrm{g}}/\rho_{\mathrm{d}}$=100. Our mass-loss rate
estimates are in agreement with the values that were derived for
AGB stars from observations and fall within the interval $\sim
10^{-8} - 10^{-5} M_{\sun}\mathrm{yr}^{-1}$ (see
\citet{Ramstedt09}). Fig.~\ref{fig:rhod} demonstrates the model
density distribution in the dust envelope. The discontinuities
coincide with the boundaries between different components of the
dust envelope.

\subsection{Verification of unambiguity}

To judge whether a two or three-shell model fits the observed data
better, we used the $F$-test to select the most appropriate model
at a significance level of $\alpha=5\%$. We could not construct a
model consisting of one or two dust shells which could reproduce
the observed SED, for no combination of parameters did we get the
$F$ value less than critical.

\begin{figure}
\centering
\includegraphics[width=\columnwidth]{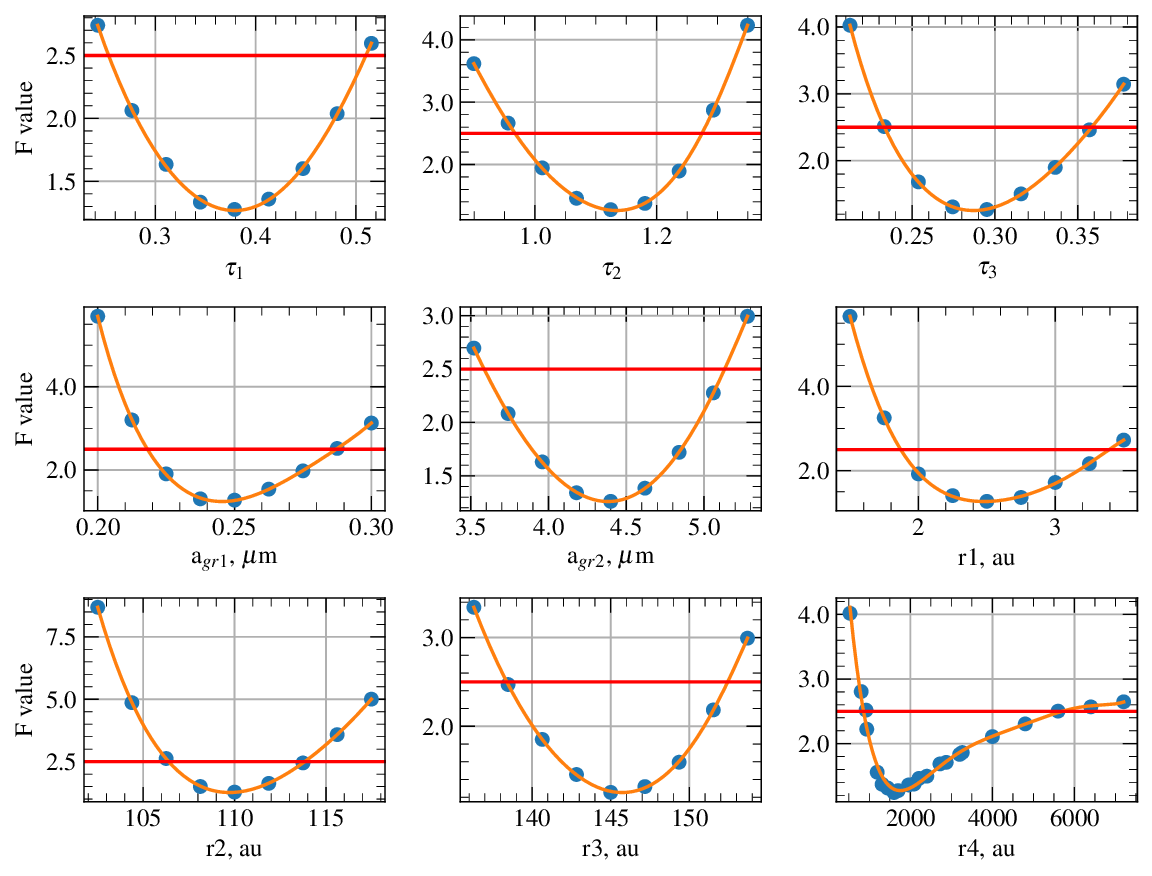}
\caption{The plot illustrating the process of deriving confidence
intervals for model parameters. Each panel shows how the $F$ value
changes with one single parameter being varied and the others
being fixed. The red line indicates the critical $F$ value.}
\label{fig:panel_pars}
\end{figure}

Our simulation is not based on analytical formulae, so, it is
rather problematic to claim the confidence intervals of the
obtained model parameters. If we consider a three-shell model,
then, in our case, for a limited sample of independent parameters
and observed data points the critical value of $F$-test statistics
is $F(5\%) = 2.5$. Our best model has $F=1.25$. We used the
$F$-test to estimate the interval for each parameter where $F <
2.5$.  So, we successively varied each of model parameters with
the others being fixed until $F$ was less than 2.5. Then we
considered the range between derived extreme values as a confident
interval for that parameter. The corresponding $F$ values for each
variation are shown in Fig.~\ref{fig:panel_pars}. The derived
intervals for all model parameters are listed in
Table~\ref{tab:params}.  One can see that the most strictly
constrained parameters are $r_2$ and $r_3$, in the sense that a
slight relative variation significantly affects the shape of SED,
whereas $r_4$, the outer radius of dust envelope, has the least
effect: even a three time increase keeps $F<2.5$. This can be
understood if we turn to the aspects of modelling: with $\tau_3$
being fixed, an increase in $r_4$ leads to a decrease in the dust
temperature at the outer radius ($r_4$) and to some redistribution
of matter inside the outer shell, so, that the dust density at its
inner radius $r_3$ declines only slightly, thus barely affecting
the shape of SED. An increase of far-IR emission produced by cold
dust at a new larger $r_4$ is partially compensated by a decrease
of dust density at the previous distance $r_4$.

\begin{figure}
\centering
\includegraphics[width=\columnwidth]{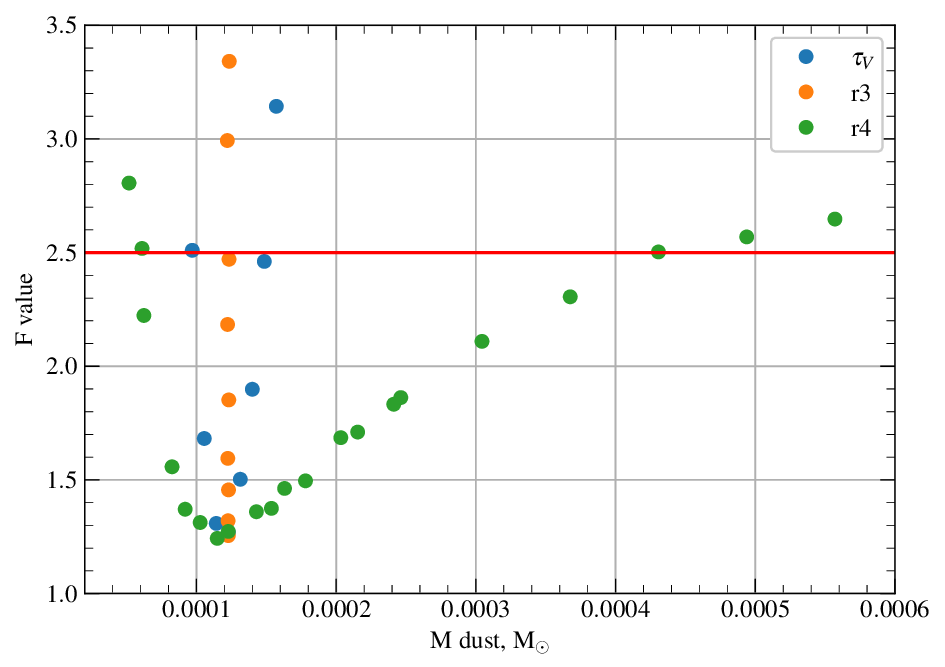}
\caption{ The plot illustrating the change of the outer shell mass
and corresponding $F$ value due to variation of one of the outer
shell parameters $\tau_{V}$, $r_3$, $r_4$ with the rest being
fixed.} \label{fig:mass3}
\end{figure}

The variation of model parameters affects the total mass of the
dust envelope differently. We calculated the total dust mass for
each model that satisfied the criterion $F<2.5$.
Fig.~\ref{fig:mass3} visualizes how the $F$ value changes with the
change of the outer dust shell mass caused by the variation of one
of model parameters, namely those corresponding to the outer
shell: $r_3$, $r_4$ -- its inner and outer radii, $\tau_{V}$. It
is clearly seen that the dust mass strongly depends only on the
outer radius. Other parameters modify the SED but barely affect
the total dust mass.

Our calculations show that the shell radius does not affect the
derived mass-loss rate, even though it does affect the resulting
mass. It is due to the fact that the lifetime of the shell and its
mass grow linearly with radius, if we assume $\rho(r)\sim r^{-2}$.


\begin{table*}
\begin{center}
\caption{The parameters of the model dust envelope components.}

\begin{tabular}{|c|c|c|c|}
\hline
& \textnumero{1} & \textnumero{2} & \textnumero{3} \\
\hline
Inner edge, au & $2.5^{+0.9}_{-0.6}$ & $110^{+4}_{-3}$ & $146^{+7}_{-7}$ \\
Outer boundary, au & $110^{+4}_{-3}$ & $146^{+7}_{-7}$ & $1750^{+3900}_{-900}$ \\
Dust grain radius, $\micron$ & $0.25^{+0.04}_{-0.03}$ & $0.25^{+0.04}_{-0.03}$ & $4.4^{+0.7}_{-0.8}$ \\
Optical depth, $\tau_V$ & $0.38^{+0.13}_{-0.13}$ & $1.12^{+0.15}_{-0.16}$ & $0.3^{+0.06}_{-0.06}$ \\
Dust mass, $M_{\sun}$ & $7^{+2.3}_{-2.2}\times 10^{-9}$ & $1.2^{+0.3}_{-0.2}\times 10^{-6}$ & $1.1^{+3}_{-0.5}\times 10^{-4}$ \\
Mass-loss rate, $M_{\sun}$yr$^{-1}$ & 1.7$\times 10^{-8}$ & 6.8$\times 10^{-6}$ & 1.1$\times 10^{-5}$ \\
\hline
\end{tabular}

\label{tab:params}
\end{center}
\end{table*}



\subsection{The necessity of large dust grains}

The main peculiarity of this model is the presence of large dust
grains with a radius of 5~$\micron$ in the third shell. The usage
of such large particles is dictated by the necessity to reproduce
the SED in the 25~$\micron$ region.

\begin{figure}
\centering
\includegraphics[width=\columnwidth]{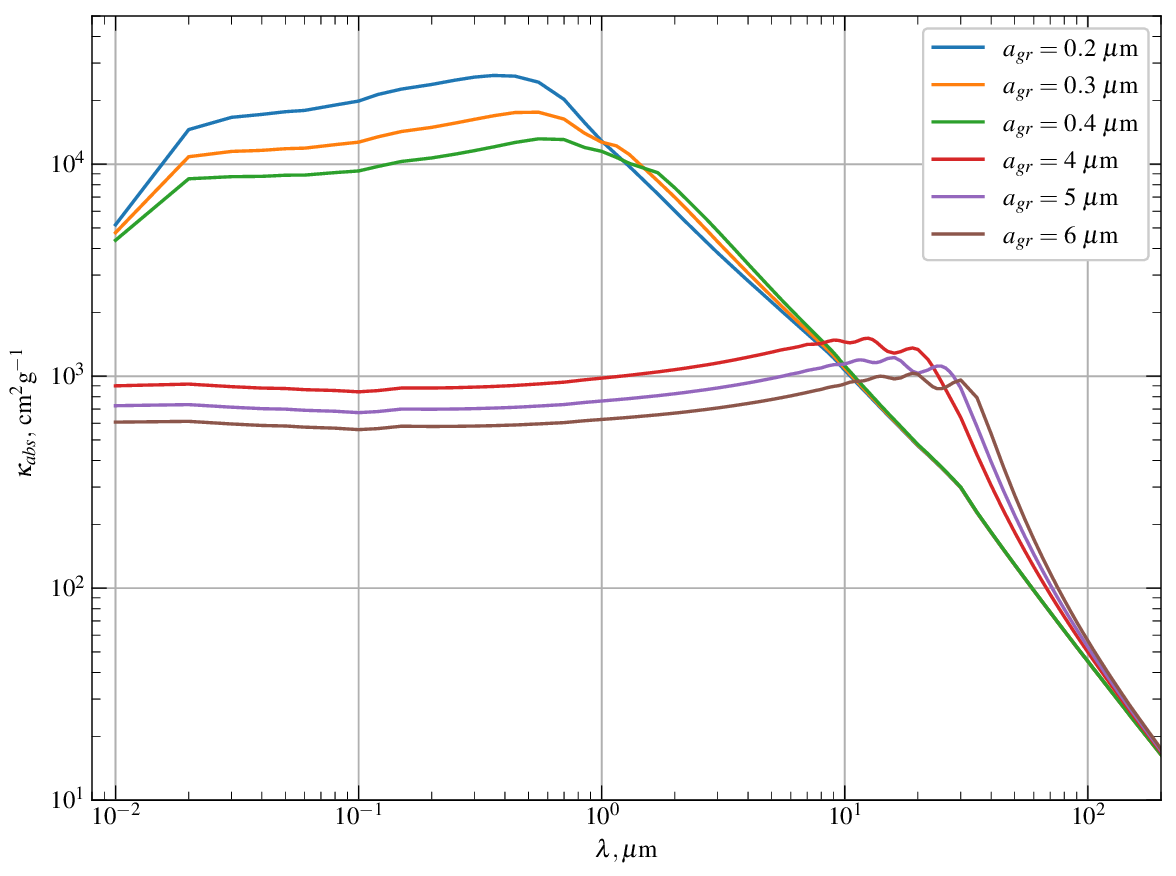}
\caption{Mass absorption coefficient $\kappa_\mathrm{abs}$ for spherical
amorphous carbon dust grains with different radii $a_\mathrm{gr}$.
The calculation was performed according to the Mie theory with
the optical constants for amorphous carbon adopted from \citet{Suh00}.} \label{fig:kappa_abs}
\end{figure}

Carbon dust grains with a radius of $\sim\!1\,\micron$ have a
larger absorption opacity coefficient at 25\,$\micron$ than the
smaller ones with $a_\mathrm{gr}\sim0.1\,\micron$.  At the same
time, the absorption opacity coefficient of the larger grains with
$a_\mathrm{gr}\sim 1\,\micron$ is much smaller in the optical
region than that of the smaller ones with
$a_\mathrm{gr}\sim0.1\,\micron$. Hence, when we try to reproduce
the SED in the 25\,$\micron$ region using the dust grains with the
radius smaller than 5\,$\micron$, a large mass of dust that is
required for this will cause too much absorption in the optical
range. But in the case of the particles with a radius of
5\,$\micron$ the values of absorption and reradiation caused by
the third dust shell component are consistent with the observed
SED. The absorption opacity coefficients for dust grains of
different sizes are depicted in Fig.~\ref{fig:kappa_abs}.


\subsection{A short discussion on the dust grain sizes}

As a result of our research two populations of dust grains were
found: smaller grains with $a_\mathrm{gr}$ = 0.3\,$\micron$ and
larger ones with $a_\mathrm{gr}$ = 5\,$\micron$.

The radius of small grains we found here coincides, in order of
magnitude, with the sizes of dust particles claimed by the recent
observational and theoretical studies. Dust grains of the similar
radii were detected by \citet{Norris12}
($a_\mathrm{gr}\sim0.3\,\micron$) for several O-rich AGB stars and
by \citet{Ohnaka17} for W~Hya ($a_\mathrm{gr}$ is about
0.1\,$\micron$ at minimum light and 0.5\,$\micron$ at maximum
light). These results are in agreement with the theoretical models
of \cite{Hofner08} who deduced that silicate dust grains of M-type
AGB stars should have radii in the range of about
0.1--1\,$\micron$ to drive the stellar outflow. In the case of
C-type AGB stars where carbon species dominate, the modelling
presented by \citet{Mattsson10} shows that the dust with
$a_\mathrm{gr}\sim 0.1$\,$\micron$ may be common for stellar
winds.

As for the large dust grains, there are some studies that have
found evidence for the presence of rather big dust particles in
the circumstellar discs around the stars more evolved than
IRAS~02143+5852. \citet{Jura97} proposed that there is an
orbiting, long-lived gravitationally bound disk of dust grains
with radii $\geq$ 0.02\,cm surrounding a carbon-rich AGB star in
the Red Rectangle nebula. \citet{Shenton95} analysed the
IRAS-millimetre flux distributions of the two post-AGB stars
AC~Her and 89~Her and suggested the existence of large grains
($a_\mathrm{gr}\geq1\,\micron$) in their dusty discs.
\citet{DeRuyter06} explored a sample of post-AGB stars and found
the emission from large ($\geq0.1$\,mm) grains. They assumed that
the presence of such large grains may be the indication of
Keplerian discs around binary systems, where the grain growth is
facilitated. Hence, according to these papers large dust grains
around IRAS~02143+5852 is a possible indication of the second
component which has not been discovered yet and the dusty disc
surrounding the possible binary system.

\section{DISCUSSION}
\label{disc}

Our photometric study has produced evidence that IRAS~02143+5852
is a pulsating variable with a period of about 25~d and light
curves typical for W~Vir stars. The star also demonstrates some
similarity to the RV~Tau stars, namely, a pattern of alternating
minima observed in the $U$ and $JHK$ phase light curves. The
double wave of alternating deep and shallow minima in the light
curve of the RV Tau variables is their main characteristic.
\citet{G1929} proposed that the recurring feature of alternating
minima represents two pulsation modes simultaneously excited in a
ratio of 2:1 resonance. A detailed description of this and other
explanations for this phenomenon is presented by
\citet{Pollard96}. It should be mentioned that much less
pronounced period doubling behaviour has recently been discovered
in some W Vir \citep{Plachy18}, BL Her \citep{Smolec12} and RR~Lyr
\citep{Szabo10} type stars.

We have found that hydrogen emission is present in the spectrum of
of IRAS~02143+5852 at certain phases of the pulsation cycle with
H$\alpha$ being the most prominent. We have also detected the
\ion{Ba}{II} and \ion{He}{I} $\lambda$10830 emissions in the
spectrum obtained close to maximum light.

\citet{schmidt04b} studied the H$\alpha$ and helium emissions in
long-period (with periods longer than 8 days) Type II Cepheids.
\citet{kovt11} detected and investigated the behaviour of emission
and line doubling of many metallic lines, in particular,
\ion{Ba}{II}, in the spectrum of the Type II Cepheid W~Vir. Strong
metallic emissions were observed in the spectra of the RV~Tau
stars U~Mon and AC~Her \citep{bopp84}.

\citet{Pollard97} reviewed the previous spectroscopic studies of
the RV~Tau variables and presented the results of a long-term
photometric and spectroscopic study of eleven stars classified as
RV~Tau. In particular, they concluded that line emission
originates within the de-excitation zone of the shock wave that
propagates through the stellar atmosphere. The fact that H$\alpha$
is observed in emission for much of the pulsation cycle is due to
a large range of atmospheric layers where this line forms. On the
contrary, the metallic emissions appear only when the shock wave
crosses the region where these lines originate. Thus,
IRAS~02143+5852 with emission lines of hydrogen, helium, and
metals in the spectrum is representative of W~Vir and RV~Tau
stars.

Significant excesses of near, medium and far-IR radiation are
distinguishing features of IRAS 02143+5852. These are due to
multiple circumstellar envelope, which according to our
simulations, has a complicated structure and was formed during
three episodes of mass loss.

Many RV~Tau stars have an IR-excess which indicates the presence
of circumstellar material and for some of them a disc-like
structure was proved. Moreover, as the discs are found around
binaries, RV~Tau stars are likely to be binary systems, too
\citep{Manick17}.

The near-IR excess is rare among W~Vir stars. Using published data
on periods and SEDs for the Galactic Type II Cepheids,
\citet{Saario18} found that only 43 objects of 1307 have a near-IR
excess and only 8 of these 43 have periods less than 30\,d.

Besides, IRAS~02143+5852 stands out among typical W~Vir stars
because of its location in the Galaxy. Whereas W~Vir stars are
low-mass pulsating variables of the intermediate disc or halo
population, IRAS~02143+5852 lies near the Galactic plane
($b=-1.^{\circ}93$).

\subsection{Objects analogous to IRAS~02143+5852}\label{analogs}

Among the W~Vir stars with the longest periods and RV~Tau stars
with the shortest periods, there are few objects similar to
IRAS~02143+5852.

{\bf CC~Lyr}. CC~Lyr is classified in the General Catalogue of
Variable Stars (GCVS) as a W~Vir-type variable with $P=24.16$\,d
and a full peak-to-peak amplitude of $\Delta V=0.8$\,mag
\citep{samus17}.

The light curves of this star, as well as those of
IRAS~02143+5852, show alternating minima of different depth, a
characteristic of RV~Tau variables \citep{schmidt04a}. CC~Lyr, as
well as IRAS~02143+5852, has a IR excess \citep{schmidt15}, also
invariably a characteristic of RV~Tau variables. CC~Lyr and
IRAS~02143+5852 have similar spectra (see Section \ref{spectra}).

\citet{Maas07} determined the parameters of the star
$T_{\text{eff}}=6250$~K and $\log g=1.0$ and revealed that
refractory elements show large depletion (e.g., [Fe/H]=--3.5).
\citet{Aoki17} confirmed its extremely low metallicity
([Fe/H]$\,<-3.5$) and concluded that the abundance anomaly of this
star is due to dust depletion. In addition, \citet{Aoki17} found
that the double-peaked H$\alpha$ emission shows no evident
velocity shift which suggests that the emission is forming in the
circumstellar matter, presumably the rotating disc around the
object. The double-peaked H$\alpha$ emission is also observed in
the spectrum of IRAS~02143+5852 at certain phases of the pulsation
cycle.

So, IRAS~02143+5852 and CC~Lyr are similar in showing alternating
minima and in having close pulsation periods and composite spectra
that combine the properties of a hot and a cool star. At the same
time, IRAS~02143+5852 demonstrates a significantly larger IR
excess than CC~Lyr. Besides, they differ in their location in the
Galaxy: whereas CC~Lyr is a high galactic latitude star
($b=+17.^{\circ}27$), IRAS~02143+5852 lies close to the Galactic
plane.

{\bf AF~Crt = IRAS 11472-0800}. According to GCVS it is a SRB-type
variable (semiregular late-type giants with poorly defined
periodicity) \citep{samus17}.

\citet{kiss07} determined a period of $P=31.5\pm0.6$~d based on
the ASAS data and classified the star as a Population II Cepheid.
Through analysing the photometric data obtained at the Valparaiso
University Observatory (VUO) in 1995--2008 \citep{VanWinckel12}
derived well-determined period values of 31.16$\pm$0.01~d ($V$)
and 32.18$\pm$0.04~d ($R$) \citep{VanWinckel12}.

We have analysed the ASAS-SN data \citep{shap14, koch17} in the
$V$-band obtained from 2012 to 2018 and determined a period of
$P=31.7$\,d and a peak-to-peak amplitude of $\Delta V=0.8$\,mag.
The light curves look like that of a W~Vir star, but the period
lies in the range typical for RV~Tau stars. Besides, the
photometric data folded on twice the period demonstrate the
alternation of more and less deep minima as is common for RV~Tau.

AF~Crt, as well as CC~Lyr, is located at high galactic latitude
($b=+51.^{\circ}56$). It is an extremely depleted object with the
photospheric abundances of [Fe/H]=--2.7 and [Sc/H]=--4.2
\citep{VanWinckel12}. \citet{VanWinckel12} also found that besides
the variation of radial velocity related to pulsations, there is a
long-term systematic change which they attributed to binary
motion.

AF~Crt is a highly evolved star of spectral type F, with a large
IR excess produced by thermal radiation of circumstellar dust.
Based on all the data known to date, \citet{VanWinckel12}
concluded that AF~Crt is a low-luminosity analogue of the dusty
RV~Tau stars.

{\bf GK~Car and GZ~Nor}. GK~Car (IRAS~11118-5726) and GZ~Nor
(IRAS~16278-5526) are two similar objects studied in detail by
\citet{Gezer19}. From the ASAS-SN data, they determined the
pulsation periods of 27.6\,d for GK~Car and 36.2\,d for GZ~Nor.
The light curves folded on twice the period demonstrate
alternating minima. The refractory elements are depleted in both
stars which may indicate their binarity. The luminosities derived
from the period-luminosity relation for the Type II Cepheids are
significantly lower than the typical values for post-AGB stars.
Based on the acquired results, \citet{Gezer19} concluded that
GK~Car and GZ~Nor are low-luminous, depleted RV~Tau stars and have
likely evolved off the RGB. GK~Car and GZ~Nor with the
temperatures of 5500 and 4875\,K, respectively, are cooler than
IRAS~02143+5852.

\section{CONCLUSIONS}\label{concl}

The results of the photometric and spectroscopic observations of
the post-AGB candidate IRAS~02143+5852 have been reported.

Based on our observational data we have found the star to vary in
brightness with a period of 25\,d and a $V$-band amplitude of
0.9\,mag. The shape of light curves is typical for the Type II
Cepheids. Its spectral properties, in particular the presence of
the strong H$\alpha$ emission at certain phases of the pulsation
cycle, are also characteristic for the Type II Cepheids.

IRAS~02143+5852 has a significant IR excess which is not usual for
the W~Vir stars, but is common among the RV~Tau variables and is a
determining feature of post-AGB objects. Modelling the SED in a
wide wavelength range 0.44--160\,$\micron$ allowed us to derive
the parameters of the dust circumstellar envelope. We show that it
has a complicated structure and was produced by three episodes of
mass loss (see Section~\ref{sed}).

Comparing IRAS~02143+5852 with similar objects leads to a
conclusion that this star may be considered as a low-luminosity
analogue of the dusty RV~Tau stars. In order to clarify its
evolutionary status it is highly necessary to obtain
high-resolution spectra which can provide abundances and stellar
parameters ($T_{\text{eff}}$, $\log g$).

Long-term high-resolution spectroscopic monitoring is also highly
desirable to derive the radial velocity curve and possibly get
evidence for the second component, as binarity is considered
related to the existence of a powerful dust envelope. As was
mentioned above, the presence of excess near-IR radiation is
characteristic of post-AGB binary stars, and those of RV~Tau type
in particular \citep{VanWinckel17}.

In addition, it is necessary to continue the photometric
monitoring to derive the so-called observed minus calculated (O-C)
diagrams to be able to detect the possible light-time travel
effect owing to the presence of a binary. Using the method
proposed by \citet{Hajdu15}, \citet{gj2017} revealed 20 new
possible binary systems among 335 Type II and anomalous Cepheids
in the Small and Large Magellanic Clouds.

\section*{ACKNOWLEDGMENTS}

The work was carried out using the equipment purchased under the
MSU Program of Development. We are indebted to the CMO staff
observers and engineers who assisted in obtaining observations. We
appreciate the comments of the referee, which led to considerable
improvement in the paper. S.Yu.~Shugarov (obtaining
$BVRI$-photometry on the 0.6-m telescope at the Star\'{a}
Lesn\'{a} Observatory) acknowledges support from the Slovak
Research and Development Agency (No. APVV-20-0148, APVV-15-0458)
and the Slovak Academy of Sciences (grant VEGA No. 2/0030/21). The
work of NPI (obtaining photometry with RC-600, analysis of the
photometric and spectroscopic data) and AVD (spectrum processing)
is supported by the RScF grant 23-12-00092. S.Zheltoukhov
(obtaining NIR photometry with ASTRONIRCAM, dust model
simulations) acknowledges the support of the Foundation for the
Development of Theoretical Physics and Mathematics BASIS (project
no. 21-2-10-35-1). The work of AMT (obtaining NIR photometry with
ASTRONIRCAM, analysis of the results of SED modelling) is
supported by the RScF grant 23-22-00182. We would like to thank
the anonymous referee for the comments that helped to sharpen the
text.

\section*{DATA AVAILABILITY}

The data underlying this study are available in the main body of the
article and in online supplementary material.

\selectlanguage{english}

\appendix
\newpage
\onecolumn
\section{$UBVR_cI_c$-observations of IRAS 02143+5852 in 2018-2021}

\begin{center}
\begin{longtable}{ccccccc}
  \caption{$UBVR_cI_c$-observations of IRAS 02143+5852 in 2018-2021}
\label{phot}\\

\hline

 JD& $U$  &$B$ & $V$ & $R_C$ & $I_C$ & Telescope\\

  \hline

  \endfirsthead

   \multicolumn{2}{l}{continued Table\ref{phot}}\\

   \hline

 JD& $U$  &$B$ & $V$ & $R_C$ & $I_C$ & Telescope\\

   \hline
   \endhead

2458133.219 &   15.830  &   15.389  &   14.160  &   13.410  &   12.609 & Sl600  \\
2458134.254 &   15.605  &   15.166  &   14.004  &   13.272  &   12.494 & Sl600  \\
2458165.308 &   15.144  &   14.762  &   13.586  &   12.865  &   12.065 & Sl600  \\
2458169.342 &   15.503  &   14.923  &   13.645  &   12.884  &   12.050 & Sl600  \\
2458175.249 &   16.143  &   15.331  &   13.987  &   13.157  &   12.294 & Sl600\\
2458182.360 &   16.160  &   15.624  &   14.350  &   13.548  &   12.729 & Sl600\\
2458186.334 &   15.146  &   14.740  &   13.606  &   12.925  &   12.194 & Sl600  \\
2458187.324 &   15.040  &   14.646  &   13.527  &   12.840  &   12.103 & Sl600  \\
2458188.278 &   15.121  &   14.658  &   13.534  &   12.842  &   12.072 & Sl600  \\
2458215.320 &   15.306  &   14.885  &   13.665  &   12.907  &   12.129 & Sl600  \\
2458218.304 &   15.459  &   14.927  &   13.656  &   12.872  &   12.069 & Sl600  \\
2458220.304 &   15.545  &   14.998  &   13.693  &   12.905  &   12.005 & Sl600  \\
2458360.539 &   15.282  &   14.932  &   13.783  &   13.103  &   12.338 & Sl600  \\
2458364.479 &   15.188  &   14.776  &   13.605  &   12.896  &   12.117 & Sl600  \\
2458378.449 &   16.443  &   15.706  &   14.307  &   13.459  &   12.590 & Sl600  \\
2458379.442 &   16.531  &   15.781  &   14.387  &   13.537  &   12.664 & Sl600  \\
2458380.436 &   16.519  &   15.811  &   14.430  &   13.583  &   12.722 & Sl600  \\
2458381.427 &   16.523  &   15.783  &   14.438  &   13.603  &   12.750 & Sl600  \\
2458391.461 &   15.354  &   14.848  &   13.649  &   12.906  &   12.096 & Sl600  \\
2458396.434 &   15.724  &   15.067  &   13.779  &   12.999  &   12.157 & Sl600  \\
2458397.479 &   15.782  &   15.159  &   13.840  &   13.051  &   12.209 & Sl600  \\
2458406.496 &   16.038  &   15.519  &   14.255  &   13.480  &   12.651 & Sl600  \\
2458407.385 &   15.949  &   15.462  &   14.244  &   13.474  &   12.662 & Sl600  \\
2458453.573 &   16.132  &   15.489  &   14.174  &   13.355  &   12.504 & Sl600  \\
2458471.329 &   15.773  &   15.105  &   13.825  &   13.038  &   12.193 & Sl600  \\
2458530.337 &   16.575  &   15.794  &   14.424  &   13.560  &   12.707 & Sl600  \\
2458685.524 &   15.402  &   14.969  &   13.836  &   13.145  &   12.390 & Sl600  \\
2458686.460 &   15.194  &   14.769  &   13.675  &   13.012  &   12.266 & Sl600  \\
2458699.526 &   16.124  &   15.410  &   14.063  &   13.235  &   12.418 & RC600  \\
2458703.536 &   16.296  &   15.723  &   14.357  &   13.529  &   12.669 & RC600  \\
2458711.514 &   15.272  &   14.675  &   13.626  &   12.948  &   12.208 & RC600  \\
2458716.494 &   15.527  &   15.004  &   13.736  &   12.980  &   12.181 & RC600  \\
2458719.401 &   15.793  &   15.065  &   13.785  &   12.991  &   12.204 & Sl600  \\
2458724.443 &   16.132  &   15.422  &   14.052  &   13.227  &   12.405 & RC600  \\
2458725.522 &   16.255  &   15.526  &   14.138  &   13.304  &   12.450 & RC600  \\
2458726.500 &   16.358  &   15.601  &   14.228  &   13.380  &   12.551 & RC600  \\
2458726.583 &   16.359  &   15.626  &   14.241  &   13.418  &   12.563 & Sl600  \\
2458727.388 &   16.359  &   15.678  &   14.290  &   13.438  &   12.619 & RC600  \\
2458728.467 &   16.395  &   15.738  &   14.360  &   13.518  &   12.676 & RC600  \\
2458729.499 &   16.474  &   15.798  &   14.433  &   13.607  &   12.767 & RC600  \\
2458730.514 &   16.441  &   15.831  &   14.465  &   13.663  &   12.780 & RC600  \\
2458732.571 &   16.281  &   15.718  &   14.430  &   13.639  &   12.836 & RC600  \\
2458735.502 &   15.293  &   14.734  &   13.682  &   13.001  &   12.268 & RC600  \\
2458738.481 &   15.264  &   14.702  &   13.607  &   12.878  &   12.140 & RC600  \\
2458739.548 &   15.379  &   14.791  &   13.653  &   12.912  &   12.146 & RC600  \\
2458740.534 &   15.491  &   14.868  &   13.686  &   12.927  &   12.147 & RC600  \\
2458741.535 &   15.527  &   14.904  &   13.701  &   12.939  &   12.133 & RC600  \\
2458742.541 &   15.568  &   14.951  &   13.713  &   12.948  &   12.135 & RC600  \\
2458744.517 &   15.736  &   15.035  &   13.767  &   12.982  &   12.174 & RC600  \\
2458745.520 &   15.756  &   15.085  &   13.808  &   13.018  &   12.197 & RC600  \\
2458746.509 &   15.827  &   15.150  &   13.870  &   13.055  &   12.235 & RC600  \\
2458749.495 &   16.055  &   15.386  &   14.048  &   13.209  &   12.388 & RC600  \\
2458750.554 &   16.133  &   15.474  &   14.132  &   13.292  &   12.466 & RC600  \\
2458758.595 &   15.507  &   15.113  &   13.996  &   13.303  &   12.548 & Sl600  \\
2458761.395 &   15.167  &   14.644  &   13.601  &   12.915  &   12.185 & RC600  \\
2458764.428 &   15.478  &   14.915  &   13.735  &   12.988  &   12.197 & RC600  \\
2458766.398 &   15.646  &   14.982  &   13.745  &   12.975  &   12.176 & RC600  \\
2458767.306 &   15.710  &   15.012  &   13.757  &   12.975  &   12.171 & RC600  \\
2458769.509 &   15.795  &   15.099  &   13.798  &   13.009  &   12.200 & RC600  \\
2458770.428 &   15.883  &   15.138  &   13.825  &   13.034  &   12.219 & RC600  \\
2458770.593 &   15.875  &   15.160  &   13.869  &   13.041  &   12.231 & Sl600  \\
2458771.402 &   16.003  &   15.194  &   13.878  &   13.045  &   12.248 & RC600  \\
2458771.566 &   16.210  &   15.226  &   13.892  &   13.105  &   12.269 & Sl600  \\
2458772.561 &   16.005  &   15.302  &   13.973  &   13.169  &   12.332 & Sl600  \\
2458773.393 &   16.104  &   15.401  &   14.028  &   13.210  &   12.383 & RC600  \\
2458774.296 &   16.268  &   15.526  &   14.122  &   13.281  &   12.451 & RC600  \\
2458774.420 &   16.270  &   15.557  &   14.165  &   13.320  &   12.470 & Sl600  \\
2458775.354 &   16.293  &   15.647  &   14.228  &   13.387  &   12.552 & RC600  \\
2458776.377 &   16.435  &   15.728  &   14.313  &   13.466  &   12.621 & RC600  \\
2458776.558 &   16.474  &   15.730  &   14.354  &   13.505  &   12.652 & Sl600  \\
2458777.411 &   16.546  &   15.806  &   14.387  &   13.549  &   12.710 & RC600  \\
2458778.345 &   16.567  &   15.872  &   14.461  &   13.615  &   12.768 & RC600  \\
2458779.396 &   16.558  &   15.907  &   14.519  &   13.679  &   12.843 & RC600  \\
2458780.344 &   16.535  &   15.906  &   14.534  &   13.710  &   12.883 & RC600  \\
2458781.348 &   16.385  &   15.761  &   14.491  &   13.676  &   12.855 & RC600  \\
2458782.436 &   16.000  &   15.468  &   14.253  &   13.485  &   12.682 & RC600  \\
2458782.494 &   15.931  &   15.464  &   14.240  &   13.479  &   12.678 & Sl600  \\
2458783.286 &   15.609  &   15.111  &   13.981  &   13.260  &   12.493 & RC600  \\
2458784.362 &   15.198  &   14.775  &   13.665  &   12.994  &   12.253 & Sl600  \\
2458784.388 &   15.284  &   14.731  &   13.677  &   12.986  &   12.250 & RC600  \\
2458785.476 &   15.159  &   14.603  &   13.559  &   12.883  &   12.140 & RC600  \\
2458786.320 &   15.151  &   14.620  &   13.556  &   12.860  &   12.129 & RC600  \\
2458790.349 &   15.469  &   14.873  &   13.680  &   12.912  &   12.124 & RC600  \\
2458791.393 &   15.536  &   14.904  &   13.691  &   12.917  &   12.122 & RC600  \\
2458792.447 &   15.643  &   14.949  &   13.701  &   12.934  &   12.133 & RC600  \\
2458794.348 &   15.785  &   15.053  &   13.782  &   12.987  &   12.178 & RC600  \\
2458795.505 &   15.864  &   15.140  &   13.841  &   13.040  &   12.221 & RC600  \\
2458796.312 &   15.962  &   15.203  &   13.889  &   13.077  &   12.251 & RC600  \\
2458797.398 &   16.026  &   15.297  &   13.964  &   13.145  &   12.320 & RC600  \\
2458798.217 &   16.094  &   15.383  &   14.033  &   13.206  &   12.373 & RC600  \\
2458799.312 &   16.225  &   15.516  &   14.140  &   13.288  &   12.467 & RC600  \\
2458800.346 &   16.197  &   15.609  &   14.222  &   13.388  &   12.550 & RC600  \\
2458802.254 &   16.413  &   15.748  &   14.386  &   13.548  &   12.701 & RC600  \\
2458803.358 &   16.436  &   15.782  &   14.452  &   13.613  &   12.786 & RC600  \\
2458804.310 &   16.376  &   15.775  &   14.470  &   13.642  &   12.825 & RC600  \\
2458805.297 &   16.288  &   15.682  &   14.437  &   13.639  &   12.827 & RC600  \\
2458806.324 &   15.910  &   15.430  &   14.282  &   13.523  &   12.741 & RC600  \\
2458807.351 &   15.603  &   15.086  &   14.015  &   13.297  &   12.546 & RC600  \\
2458808.256 &   15.320  &   14.813  &   13.777  &   13.085  &   12.350 & RC600  \\
2458810.232 &   15.111  &   14.582  &   13.539  &   12.856  &   12.124 & RC600  \\
2458811.494 &   15.167  &   14.665  &   13.575  &   12.872  &   12.121 & RC600  \\
2458812.343 &   15.277  &   14.734  &   13.604  &   12.884  &   12.114 & RC600  \\
2458813.392 &   15.409  &   14.826  &   13.649  &   12.900  &   12.114 & RC600  \\
2458816.480 &   15.602  &   14.979  &   13.714  &   12.931  &   12.123 & RC600  \\
2458818.382 &   15.730  &   15.039  &   13.760  &   12.968  &   12.151 & RC600  \\
2458820.397 &   15.849  &   15.148  &   13.840  &   13.025  &   12.205 & RC600  \\
2458822.393 &   16.053  &   15.322  &   13.963  &   13.142  &   12.314 & RC600  \\
2458823.373 &   16.195  &   15.420  &   14.043  &   13.208  &   12.379 & RC600  \\
2458824.420 &   16.228  &   15.520  &   14.129  &   13.287  &   12.460 & RC600  \\
2458825.381 &   16.308  &   15.601  &   14.204  &   13.373  &   12.532 & RC600  \\
2458826.381 &   16.370  &   15.678  &   14.282  &   13.454  &   12.612 & RC600  \\
2458827.377 &   16.438  &   15.730  &   14.362  &   13.520  &   12.689 & RC600  \\
2458828.370 &   16.410  &   15.761  &   14.401  &   13.572  &   12.738 & RC600  \\
2458829.406 &   16.552  &   15.749  &   14.405  &   13.597  &   12.771 & RC600  \\
2458830.386 &   16.174  &   15.632  &   14.352  &   13.545  &   12.741 & RC600  \\
2458833.363 &   15.357  &   14.841  &   13.764  &   13.063  &   12.314 & RC600  \\
2458834.302 &   15.224  &   14.695  &   13.640  &   12.951  &   12.204 & RC600  \\
2458835.306 &   15.182  &   14.663  &   13.598  &   12.901  &   12.160 & RC600  \\
2458836.289 &   15.209  &   14.699  &   13.615  &   12.909  &   12.156 & RC600  \\
2458837.263 &   15.308  &   14.750  &   13.652  &   12.924  &   12.150 & RC600  \\
2458838.277 &   15.342  &   14.811  &   13.666  &   12.926  &   12.151 & RC600  \\
2458840.352 &   15.474  &   14.892  &   13.676  &   12.918  &   12.131 & RC600  \\
2458841.376 &   15.543  &   14.932  &   13.695  &   12.929  &   12.134 & RC600  \\
2458843.314 &   15.645  &   15.012  &   13.745  &   12.948  &   12.154 & RC600  \\
2458844.294 &   15.702  &   15.046  &   13.780  &   12.991  &   12.182 & RC600  \\
2458846.340 &   15.849  &   15.185  &   13.877  &   13.075  &   12.249 & RC600  \\
2458847.403 &   15.936  &   15.275  &   13.950  &   13.137  &   12.317 & RC600  \\
2458847.475 &   15.909  &   15.299  &   13.976  &   13.179  &   12.325 & Sl600  \\
2458848.277 &   15.984  &   15.344  &   14.013  &   13.193  &   12.371 & RC600  \\
2458849.240 &   16.008  &   15.406  &   14.083  &   13.259  &   12.431 & RC600  \\
2458851.271 &   16.143  &   15.518  &   14.203  &   13.396  &   12.571 & RC600  \\
2458852.236 &   16.146  &   15.541  &   14.252  &   13.438  &   12.631 & RC600  \\
2458853.272 &   16.101  &   15.550  &   14.301  &   13.504  &   12.696 & RC600  \\
2458854.436 &   16.064  &   15.550  &   14.319  &   13.523  &   12.723 & RC600  \\
2458855.224 &   15.988  &   15.506  &   14.292  &   13.513  &   12.721 & RC600  \\
2458855.263 &   15.930  &   15.493  &   14.252  &   13.510  &   12.711 & Sl600  \\
2458856.339 &   15.747  &   15.278  &   14.095  &   13.384  &   12.590 & Sl600  \\
2458857.266 &   15.582  &   15.026  &   13.958  &   13.225  &   12.477 & RC600  \\
2458859.215 &   15.213  &   14.650  &   13.599  &   12.915  &   12.178 & Sl600  \\
2458859.262 &   15.115  &   14.682  &   13.562  &   12.899  &   12.161 & Sl600  \\
2458862.221 &   15.377  &   14.817  &   13.672  &   12.936  &   12.160 & RC600  \\
2458863.217 &   15.472  &   14.876  &   13.698  &   12.947  &   12.157 & RC600  \\
2458863.299 &   15.405  &   14.907  &   13.683  &   12.946  &   12.136 & Sl600  \\
2458865.229 &   15.571  &   14.940  &   13.700  &   12.931  &   12.131 & RC600  \\
2458866.238 &   15.649  &   14.972  &   13.709  &   12.925  &   12.133 & Sl600  \\
2458867.199 &   15.698  &   14.999  &   13.725  &   12.949  &   12.135 & RC600  \\
2458867.241 &   15.750  &   15.032  &   13.743  &   12.972  &   12.151 & Sl600  \\
2458868.227 &   15.737  &   15.033  &   13.751  &   12.977  &   12.152 & RC600  \\
2458869.252 &   15.806  &   15.084  &   13.787  &   12.988  &   12.179 & RC600  \\
2458870.176 &   15.838  &   15.146  &   13.827  &   13.024  &   12.210 & RC600  \\
2458871.214 &   15.934  &   15.226  &   13.909  &   13.089  &   12.265 & RC600  \\
2458873.199 &   16.155  &   15.497  &   14.102  &   13.253  &   12.418 & RC600  \\
2458874.360 &   16.239  &   15.610  &   14.195  &   13.346  &   12.506 & RC600  \\
2458876.325 &   16.371  &   15.767  &   14.329  &   13.482  &   12.641 & RC600  \\
2458877.190 &   16.394  &   15.794  &   14.389  &   13.543  &   12.713 & RC600  \\
2458878.193 &   16.411  &   15.837  &   14.435  &   13.609  &   12.775 & RC600  \\
2458881.211 &   15.890  &   15.387  &   14.197  &   13.433  &   12.646 & RC600  \\
2458883.307 &   15.131  &   14.645  &   13.607  &   12.929  &   12.190 & RC600  \\
2458884.203 &   15.030  &   14.522  &   13.496  &   12.831  &   12.093 & RC600  \\
2458887.165 &   15.276  &   14.694  &   13.563  &   12.847  &   12.074 & RC600  \\
2458889.309 &   15.439  &   14.814  &   13.628  &   12.858  &   12.073 & RC600  \\
2458890.186 &   15.483  &   14.855  &   13.640  &   12.872  &   12.074 & RC600  \\
2458891.189 &   15.541  &   14.897  &   13.663  &   12.889  &   12.078 & RC600  \\
2458892.224 &   15.590  &   14.959  &   13.700  &   12.924  &   12.111 & RC600  \\
2458894.207 &   15.743  &   15.082  &   13.785  &   12.978  &   12.163 & RC600  \\
2458895.287 &   15.808  &   15.179  &   13.839  &   13.042  &   12.224 & RC600  \\
2458896.253 &   15.934  &   15.264  &   13.909  &   13.096  &   12.261 & RC600  \\
2458897.266 &   15.981  &   15.357  &   13.981  &   13.163  &   12.321 & RC600  \\
2458898.211 &   16.065  &   15.436  &   14.054  &   13.225  &   12.402 & RC600  \\
2458899.251 &   16.087  &   15.511  &   14.115  &   13.293  &   12.471 & RC600  \\
2458900.361 &   16.142  &   15.585  &   14.203  &   13.377  &   12.562 & RC600  \\
2458901.328 &   16.122  &   15.640  &   14.272  &   13.448  &   12.627 & RC600  \\
2458902.300 &   16.244  &   15.657  &   14.357  &   13.479  &   12.678 & RC600  \\
2458903.305 &   16.173  &   15.652  &   14.349  &   13.541  &   12.714 & RC600  \\
2458905.216 &   15.892  &   15.407  &   14.248  &   13.479  &   12.690 & RC600  \\
2458906.229 &   15.490  &   15.065  &   13.979  &   13.256  &   12.492 & RC600  \\
2458908.237 &   14.979  &   14.508  &   13.496  &   12.836  &   12.122 & RC600  \\
2458909.203 &   15.020  &   14.514  &   13.483  &   12.813  &   12.083 & RC600  \\
2458911.223 &   15.244  &   14.720  &   13.597  &   12.870  &   12.091 & RC600  \\
2458912.208 &   15.338  &   14.801  &   13.628  &   12.877  &   12.088 & RC600  \\
2458913.210 &   15.408  &   14.844  &   13.645  &   12.882  &   12.082 & RC600  \\
2458914.212 &   15.474  &   14.887  &   13.657  &   12.881  &   12.075 & RC600  \\
2458917.194 &   15.616  &   14.986  &   13.702  &   12.906  &   12.093 & RC600  \\
2458918.232 &   15.666  &   15.028  &   13.738  &   12.935  &   12.126 & RC600  \\
2458919.209 &   15.793  &   15.099  &   13.779  &   12.979  &   12.159 & RC600  \\
2458922.246 &   16.006  &   15.430  &   14.032  &   13.187  &   12.358 & RC600  \\
2459050.529 &   16.166  &   15.581  &   14.262  &   13.426  &   12.606 & RC600  \\
2459051.519 &   16.217  &   15.642  &   14.313  &   13.492  &   12.666 & RC600  \\
2459055.528 &   15.378  &   14.930  &   13.872  &   13.147  &   12.421 & RC600  \\
2459061.510 &   15.257  &   14.767  &   13.603  &   12.853  &   12.053 & RC600  \\
2459067.498 &   15.725  &   15.028  &   13.725  &   12.926  &   12.107 & RC600  \\
2459068.456 &   15.782  &   15.094  &   13.780  &   12.971  &   12.143 & RC600  \\
2459070.469 &   15.995  &   15.301  &   13.927  &   13.092  &   12.256 & RC600  \\
2459079.550 &   15.810  &   15.283  &   14.102  &   13.338  &   12.551 & RC600  \\
2459081.448 &   15.243  &   14.725  &   13.660  &   12.953  &   12.220 & RC600  \\
2459082.472 &   15.110  &   14.570  &   13.536  &   12.838  &   12.106 & RC600  \\
2459084.503 &   15.130  &   14.605  &   13.525  &   12.820  &   12.064 & RC600  \\
2459087.396 &   15.321  &   14.762  &   13.585  &   12.833  &   12.036 & RC600  \\
2459088.327 &   15.383  &   14.818  &   13.609  &   12.841  &   12.035 & RC600  \\
2459090.527 &   15.554  &   14.903  &   13.653  &   12.871  &   12.058 & RC600  \\
2459091.545 &   15.617  &   14.972  &   13.693  &   12.899  &   12.082 & RC600  \\
2459092.499 &   15.719  &   15.039  &   13.736  &   12.936  &   12.115 & RC600  \\
2459093.450 &   15.784  &   15.108  &   13.792  &   12.982  &   12.157 & RC600  \\
2459094.488 &   15.874  &   15.196  &   13.855  &   13.046  &   12.208 & RC600  \\
2459095.447 &   15.918  &   15.275  &   13.923  &   13.104  &   12.265 & RC600  \\
2459097.488 &   16.030  &   15.420  &   14.067  &   13.240  &   12.404 & RC600  \\
2459099.437 &   16.074  &   15.499  &   14.179  &   13.390  &   12.544 & RC600  \\
2459103.417 &   15.780  &   15.312  &   14.159  &   13.390  &   12.610 & RC600  \\
2459104.492 &   15.568  &   15.093  &   13.986  &   13.251  &   12.479 & RC600  \\
2459106.506 &   15.168  &   14.651  &   13.603  &   12.910  &   12.170 & RC600  \\
2459108.568 &   15.151  &   14.601  &   13.528  &   12.818  &   12.070 & RC600  \\
2459109.364 &   15.202  &   14.684  &   13.566  &   12.833  &   12.079 & RC600  \\
2459110.325 &   15.350  &   14.766  &   13.605  &   12.860  &   12.094 & RC600  \\
2459112.515 &   15.509  &   14.883  &   13.650  &   12.879  &   12.080 & RC600  \\
2459113.278 &   15.584  &   14.932  &   13.675  &   12.879  &   12.084 & RC600  \\
2459114.517 &   15.605  &   14.932  &   13.675  &   12.892  &   12.086 & RC600  \\
2459115.260 &   15.673  &   14.971  &   13.685  &   12.901  &   12.097 & RC600  \\
2459117.464 &   15.821  &   15.062  &   13.752  &   12.957  &   12.153 & RC600  \\
2459118.287 &   15.813  &   15.110  &   13.798  &   12.998  &   12.181 & RC600  \\
2459119.312 &   15.929  &   15.208  &   13.869  &   13.052  &   12.247 & RC600  \\
2459120.381 &   16.086  &   15.323  &   13.953  &   13.136  &   12.302 & RC600  \\
2459121.407 &   16.104  &   15.455  &   14.062  &   13.211  &   12.382 & RC600  \\
2459126.457 &   16.296  &   15.711  &   14.382  &   13.532  &   12.721 & RC600  \\
2459129.448 &   15.919  &   15.339  &   14.154  &   13.374  &   12.589 & RC600  \\
2459130.487 &   15.519  &   14.972  &   13.858  &   13.132  &   12.372 & RC600  \\
2459131.303 &   15.210  &   14.693  &   13.642  &   12.950  &   12.217 & RC600  \\
2459132.374 &   15.070  &   14.529  &   13.483  &   12.813  &   12.090 & RC600  \\
2459136.461 &   15.262  &   14.696  &   13.538  &   12.793  &   12.018 & RC600  \\
2459138.490 &   15.396  &   14.807  &   13.594  &   12.819  &   12.020 & RC600  \\
2459140.460 &   15.557  &   14.904  &   13.645  &   12.864  &   12.048 & RC600  \\
2459145.560 &   15.935  &   15.272  &   13.930  &   13.114  &   12.274 & RC600  \\
2459146.445 &   15.992  &   15.334  &   13.994  &   13.178  &   12.337 & RC600  \\
2459149.469 &   16.157  &   15.543  &   14.223  &   13.391  &   12.578 & RC600  \\
2459150.524 &   16.123  &   15.558  &   14.262  &   13.441  &   12.621 & RC600  \\
2459152.451 &   16.043  &   15.440  &   14.233  &   13.450  &   12.664 & RC600  \\
2459153.440 &   15.750  &   15.263  &   14.111  &   13.387  &   12.592 & RC600  \\
2459158.447 &   15.204  &   14.652  &   13.564  &   12.852  &   12.099 & RC600  \\
2459162.502 &   15.467  &   14.859  &   13.631  &   12.865  &   12.071 & RC600  \\
2459163.436 &   15.518  &   14.885  &   13.646  &   12.864  &   12.075 & RC600  \\
2459164.382 &   15.527  &   14.918  &   13.661  &   12.879  &   12.076 & RC600  \\
2459165.385 &   15.614  &   14.947  &   13.677  &   12.893  &   12.088 & RC600  \\
2459166.307 &   15.699  &   14.984  &   13.709  &   12.921  &   12.106 & RC600  \\
2459167.389 &   15.752  &   15.042  &   13.748  &   12.953  &   12.141 & RC600  \\
2459175.480 &   16.394  &   15.646  &   14.273  &   13.440  &   12.612 & RC600  \\
2459177.416 &   16.313  &   15.684  &   14.343  &   13.517  &   12.711 & RC600  \\
2459181.399 &   15.353  &   14.793  &   13.703  &   13.000  &   12.266 & RC600  \\
2459182.359 &   15.200  &   14.624  &   13.566  &   12.875  &   12.135 & RC600  \\
2459183.445 &   15.147  &   14.590  &   13.525  &   12.833  &   12.077 & RC600  \\
2459196.309 &   15.957  &   15.344  &   13.987  &   13.172  &   12.343 & RC600  \\
2459251.359 &   16.019  &   15.601  &   14.344  &   13.560  &   12.766 & RC600  \\

  \hline

\end{longtable}
\end{center}

\section{$JHK$-observations of IRAS~02143+5852 in 2017--2021}

\begin{center}
\begin{longtable}{ccccccc}
  \caption{$JHK$-observations of IRAS~02143+5852 in 2017--2021.}
  \label{nir}\\

\hline

 JD & $J$(MKO) & $H$(MKO) & $K$(MKO) & $J$(2MASS) & $H$(2MASS) & $K$(2MASS)\\

  \hline

  \endfirsthead

   \multicolumn{2}{l}{continued Table\ref{nir}}\\

   \hline

 JD & $J$(MKO) & $H$(MKO) & $K$(MKO) & $J$(2MASS) & $H$(2MASS) & $K$(2MASS)\\

   \hline
   \endhead

2458064.331&    10.644& 9.620&  8.482&  10.809& 9.578&  8.571\\
2458089.324&    10.614& 9.534&  8.366&  10.785& 9.491&  8.457\\
2458099.152&    10.753& 9.759&  8.657&  10.912& 9.719&  8.743\\
2458144.318&    10.502& 9.494&  8.366&  10.665& 9.453&  8.454\\
2458147.243&    10.608& 9.632&  8.521&  10.767& 9.592&  8.608\\
2458151.360&    10.831& 9.839&  8.731&  10.991& 9.799&  8.818\\
2458153.394&    10.974& 9.976&  8.873&  11.134& 9.936&  8.959\\
2458156.259&    11.157& 10.129& 8.970&  11.324& 10.086& 9.060\\
2458157.168&    11.177& 10.121& 8.947&  11.347& 10.077& 9.038\\
2458158.187&    11.121& 10.048& 8.852&  11.294& 10.003& 8.945\\
2458166.282&    10.553& 9.535&  8.388&  10.718& 9.493&  8.477\\
2458172.339&    10.629& 9.663&  8.533&  10.788& 9.622&  8.621\\
2458180.208&    11.193& 10.259& 9.224&  11.343& 10.222& 9.305\\
2458480.497&    11.172& 10.207& 9.142&  11.326& 10.169& 9.226\\
2458481.238&    11.214& 10.250& 9.168&  11.370& 10.211& 9.253\\
2458482.356&    11.278& 10.307& 9.227&  11.434& 10.268& 9.312\\
2458483.283&    11.299& 10.308& 9.199&  11.459& 10.268& 9.286\\
2458484.294&    11.279& 10.273& 9.134&  11.442& 10.231& 9.223\\
2458486.385&    11.035& 9.919&  8.665&  11.216& 9.872&  8.762\\
2458487.340&    10.844& 9.687&  8.380&  11.032& 9.637&  8.481\\
2458489.326&    10.645& 9.492&  8.207&  10.831& 9.443&  8.306\\
2458493.380&    10.657& 9.597&  8.421&  10.827& 9.553&  8.512\\
2458511.491&    10.988& 9.849&  8.550&  11.174& 9.799&  8.650\\
2458513.377&    10.733& 9.573&  8.314&  10.918& 9.525&  8.411\\
2458520.406&    10.666& 9.648&  8.512&  10.830& 9.606&  8.601\\
2458862.309&    10.652& 9.598&  8.436&  10.821& 9.555&  8.527\\
2458866.295&    10.611& 9.640&  8.524&  10.770& 9.599&  8.611\\
2458867.256&    10.638& 9.661&  8.559&  10.796& 9.621&  8.645\\
2458868.292&    10.640& 9.694&  8.599&  10.795& 9.654&  8.685\\
2458869.381&    10.685& 9.739&  8.650&  10.840& 9.700&  8.735\\
2458870.398&    10.732& 9.789&  8.718&  10.885& 9.751&  8.802\\
2458871.173&    10.784& 9.828&  8.771&  10.937& 9.790&  8.854\\
2458877.203&    11.234& 10.286& 9.274&  11.383& 10.250& 9.354\\
2458879.174&    11.377& 10.385& 9.334&  11.532& 10.348& 9.417\\
2458884.287&    10.605& 9.494&  8.240&  10.785& 9.447&  8.337\\
2458908.162&    10.639& 9.490&  8.211&  10.824& 9.441&  8.310\\
2458911.153&    10.609& 9.585&  8.436&  10.774& 9.543&  8.526\\
2458912.175&    10.598& 9.596&  8.481&  10.759& 9.555&  8.568\\
2458914.159&    10.572& 9.587&  8.494&  10.730& 9.547&  8.580\\
2458916.191&    10.579& 9.612&  8.513&  10.736& 9.572&  8.599\\
2458918.161&    10.621& 9.673&  8.593&  10.775& 9.634&  8.678\\
2458919.171&    10.663& 9.710&  8.648&  10.816& 9.672&  8.731\\
2458920.193&    10.724& 9.778&  8.724&  10.876& 9.740&  8.807\\
2459084.503&    10.559& 9.463&  8.251&  10.735& 9.418&  8.345\\
2459091.541&    10.579& 9.601&  8.483&  10.738& 9.560&  8.570\\
2459095.448&    10.776& 9.813&  8.735&  10.931& 9.774&  8.820\\
2459141.444&    10.568& 9.579&  8.449&  10.730& 9.538&  8.537\\
2459149.469&    11.058& 10.072& 8.976&  11.216& 10.032& 9.062\\
2459153.340&    11.101& 10.030& 8.797&  11.276& 9.984&  8.893\\
2459155.362&    10.843& 9.708&  8.411&  11.028& 9.659&  8.511\\
2459161.566&    10.571& 9.523&  8.347&  10.740& 9.480&  8.438\\
2459165.399&    10.577& 9.573&  8.432&  10.744& 9.529&  8.521\\
2459171.457&    10.851& 9.890&  8.805&  11.007& 9.851&  8.890\\
2459175.491&    11.134& 10.193& 9.116&  11.288& 10.154& 9.200\\
2459194.435&    10.714& 9.750&  8.629&  10.872& 9.709&  8.716\\
2459195.356&    10.775& 9.802&  8.705&  10.933& 9.762&  8.791\\
2459196.297&    10.838& 9.871&  8.781&  10.994& 9.831&  8.866\\
2459199.299&    11.070& 10.092& 8.995&  11.228& 10.053& 9.081\\
2459201.303&    11.186& 10.207& 9.096&  11.345& 10.167& 9.183\\
2459206.213&    10.697& 9.529&  8.221&  10.887& 9.479&  8.322\\
2459208.145&    10.614& 9.479&  8.209&  10.798& 9.431&  8.307\\
2459209.141&    10.611& 9.522&  8.301&  10.787& 9.476&  8.396\\
2459211.209&    10.599& 9.548&  8.371&  10.769& 9.504&  8.463\\
2459220.222&    10.846& 9.888&  8.808&  11.001& 9.849&  8.892\\
2459224.299&    11.169& 10.237& 9.184&  11.320& 10.200& 9.267\\
2459239.410&    10.583& 9.585&  8.425&  10.748& 9.542&  8.515\\
2459245.151&    10.848& 9.866&  8.761&  11.007& 9.825&  8.848\\
2459246.260&    10.929& 9.956&  8.865&  11.086& 9.917&  8.950\\
2459255.160&    10.814& 9.659&  8.348&  11.003& 9.608&  8.449\\
2459257.214&    10.636& 9.515&  8.248&  10.818& 9.467&  8.346\\
2459258.143&    10.623& 9.530&  8.292&  10.801& 9.483&  8.388\\
2459264.350&    10.603& 9.597&  8.441&  10.768& 9.554&  8.531\\

\hline

\end{longtable}
\end{center}

\label{lastpage}
\end{document}